# The Augmented Potential Method: Multiscale Modeling Toward a Spectral Defect Genome

Nutth Tuchinda[1], Changle Li[1], Christopher A. Schuh[1,2*]

[1]*Department of Materials Science and Engineering, Massachusetts Institute of Technology, 77 Massachusetts Avenue, Cambridge, MA, 02139, USA*

[2] *Department of Materials Science and Engineering, Northwestern University, 2145 Sheridan Road, Evanston, IL 60208, USA*

*Correspondence to: schuh@northwestern.edu

## ABSTRACT

Modeling of solute chemistry at low-symmetry defects in materials is historically challenging, due to the computation cost required to evaluate thermodynamic properties from first principles. Here, we offer a hybrid multiscale approach called the *augmented potential method* that connects the chemical flexibility and high accuracy of a universal machine learning potential at the site of the defect, with the computational speed of an efficient potential implemented away from the defect site. The method allows us to rapidly compute distributions of grain boundary segregation energy for 1,036 binary alloy pairs (including Ag, Al, Au, Cr, Cu, Fe, Mo, Nb, Ni, Pd, Pt, Ta and V, W solvent), creating a database ~5x larger than previously published spectral compilations, and yet has improved accuracy. The approach can also address problems such as the solute-solute interactions in polycrystals that require significant computational efforts, paving a pathway toward a complete defect genome in crystalline materials.

## GRAPHICAL ABSTRACT

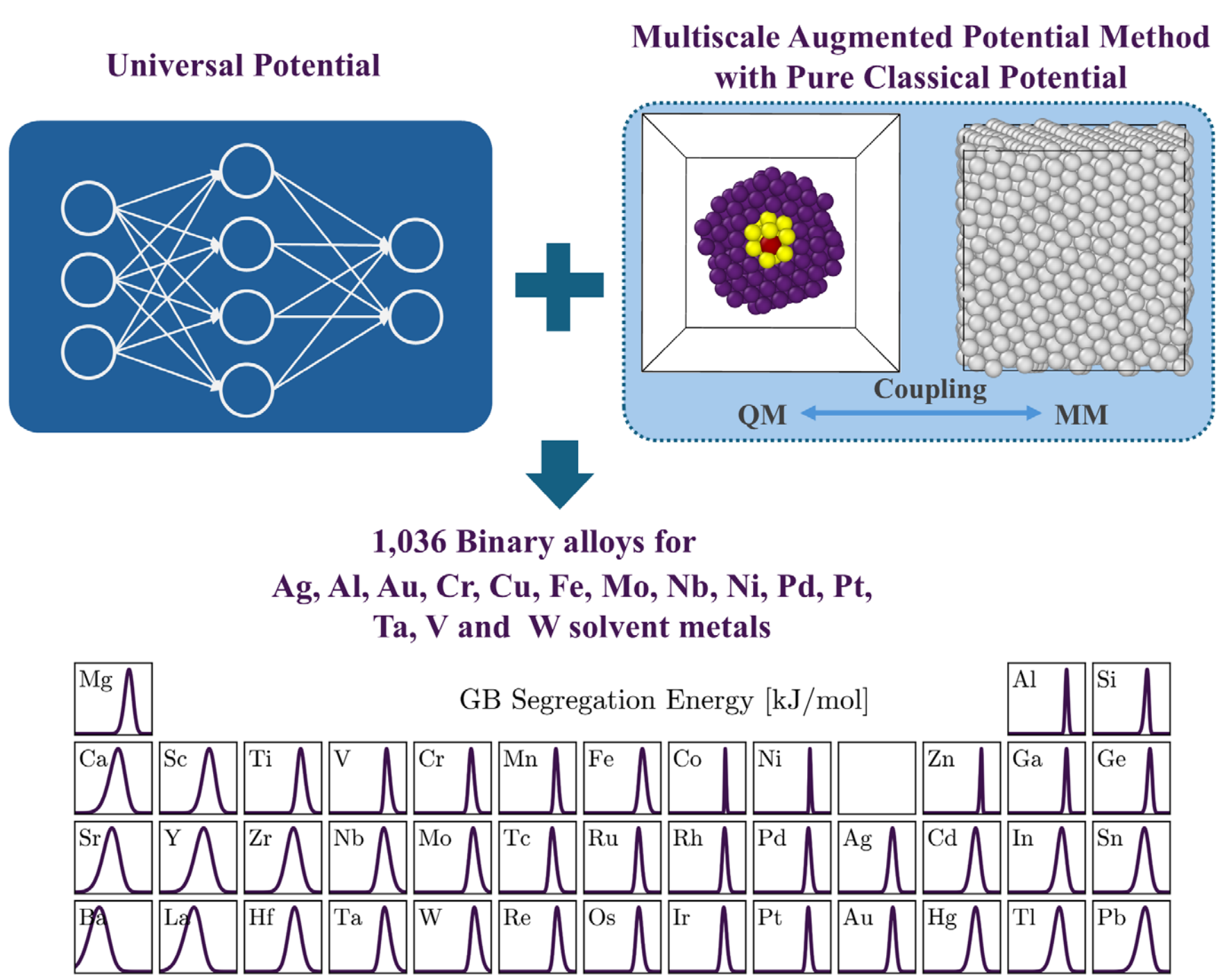

In crystalline materials, defects often control the mechanical, thermodynamic, kinetic and functional properties [1–4]. At the same time, defects are often prime segregation sites for chemical solutes incorporated either by design or incidentally as impurities, and solute segregation itself can also dramatically influence properties [5–7]. Understanding and mastering solute segregation is therefore a primary direction for the control of microstructure in materials, but it represents a longstanding problem due to its structural complexity and computational effort needed to compute segregation energy spectra for the diverse range of atomic configurations that comprise defects [8,9]. For example, polycrystalline materials comprise a spectrum of grain boundaries (GBs) and many millions of unique atomic sites associated with those defects. Addressing the segregation of solutes to those GBs requires large atomistic simulations for even a single alloy.

With the current state-of-the-art GB models, the number of atomic site computations required to estimate a binary alloy segregation spectrum has been significantly reduced [10–12] using a data-science approach. By using dimensionality reduction methods, it is possible to define a set of atomic environments that are representative of the larger problem, and then focus more quantitatively accurate computations on those sites. Given that those sites have low symmetry, they are not generally amenable to full simulation by the most accurate quantum-mechanical methods. Therefore, a recent focus has been on the implementation of multiscale methods, which treat the neighborhood of the solute segregation with quantitative accuracy, while connecting to faster methods at far distances where accuracy is not needed [9,13–20]. As an example, a hybrid quantum mechanical-molecular mechanics method allowed spectral evaluations of binary Al-based alloys without any reliance on binary interatomic potentials [9]. Nevertheless, quantum calculations remain challenging for high-throughput calculations [21,22] of large polycrystalline systems even with the current computational resources. For example, such simulations cannot yet efficiently address systems with magnetic effects, nor can they handle the full problem of solute segregation beyond the dilute limit (where solutes interact with one another in nearby defect sites) due to the computational intensity.

In this work, we develop an "augmented potential method", coupling a fast classical interatomic potential to a computationally intensive universal potential. This approach simultaneously provides total chemical flexibility and extreme computational speed while maintaining state-of-the-art accuracy vis-à-vis classical potentials. The accuracy of the method is established against full quantum simulations for a sample of target sites, and the speed of the method is established by developing a complete evaluation of 1,036 binary alloys based on solvents of Ag, Al, Au, Cr, Cu, Fe, Mo, Nb, Ni, Pd, Pt, Ta, V and W and 75 solutes from Li to Bi. Finally, the treatment of bcc Fe and the computation of solute-solute interactions provide demonstrations of computational efficiency that is orders of magnitude beyond the state of the art with first-principles calculations. The result is a comprehensive defect spectrum atlas for use with future alloy design and optimization.

The method is illustrated in Fig. 1. We follow Refs. [9–12] to generate polycrystals and relax them using molecular dynamics and statics in Fig. 1a. To address the full spectrum of GBs sites in random polycrystals, our model contains 12 grains previously shown to be representative of random polycrystals [23–25]. The random polycrystalline seeds and grain sizes used here are well-established as providing good representative polycrystal results without spurious sensitivity [23,26,27]. The site sampling (Fig. 1b-d) is done using a data science approach using a Python workflow [28–39]. In short, local atomic environment descriptors in Fig. 1b are calculated using the Smooth Overlap of Atomic Positions (SOAP) method [40–42], followed by dimensionality reduction [43–45] in Fig. 1c, then a site sampling algorithm [46,47] in Fig. 1d from the reduced GB configurational space. We apply a cutoff radius summarized in the Supplemental Material [48], along with a Gaussian smearing width of 0.5 Å and maximum radial ($n$) and angular ($l$) bases of 10 and 5 respectively. This results in a SOAP vector of length 331 for each atomic site, of which their principal components can be computed; we narrow to the 1-10$^{th}$ highest variance components, which have been shown to represent GB local atomic environments well [9]. Lastly, we sample GB sites from the reduced GB configurational space with a k-means clustering algorithm [46,47] and choose representative GB sites from 100 centroids on the reduced GB space. In essence, the algorithm uses the SOAP vectors of the sample sites to train and predict the segregation energies of all sites in the polycrystals for skew-normal distribution fitting; the learning algorithm effectively weights the sampled sites to produce the correct distribution present in a polycrystal. More details can be found in Supplemental Material S1 [48].

The hybrid setup for typical multi-scale molecular simulations is shown in Fig. 1e-f where the system is classified into a core, buffer and molecular mechanics (MM) region [14,16,17]. Conventionally for this kind of approach, the core and buffer region are computed from quantum mechanical calculations to overcome the chemical limitations of the large-scale interatomic potential used for the MM region. However, for many metallic systems, the buffer size needed is still rather large compared to typical first principles calculations. For example, using force convergence at a bulk Ni site, the total number of atoms in the spherical zone is ~200, which is a demanding quantum calculation. Once the scope is expanded to consider low-symmetry GB sites, solute-solute interactions, and many possible alloys, the combinatorial growth of the computation time becomes truly prohibitive for quantum-based computations.

Recent computational advances in the area of universal machine learning interatomic potentials [49–57] are very promising for multiscale challenges like this one, providing an attractive speed-accuracy tradeoff to bypass expensive ab initio computations. For example, the EquiformerV2 model [49,52,58,59] provides a promising candidate for describing chemistry for multiple inorganic databases [60–62]. Our approach is therefore to replace the QM calculations in the core and buffer region of a multiscale model, with a universal machine learning potential [52,54,63–66] for computing solute energetics at defect sites. This results in a multifold reduction in computation time, especially with graphical processing unit acceleration, yet can retain accuracy close to quantum approaches. The force convergence of the buffer zone in such an implementation is shown in Supplemental Material S2 for all MM potentials used here, and the MM region is computed via the LAMMPS software package [67–70]).We term this approach the Augmented Potential Method (APM), since it uses a more elaborate and accurate potential in combination with a rapid one.

We show a validation of APM vis-à-vis QM calculation with a density functional theory (DFT) method from Ref. [71] in Fig. 2. This validation considers all three unique sites in the Σ5[001](210) GB, with Al as the base metal and a wide range of solute substitutions. The dilute-limit segregation energy at site $i$ is defined by the energy change upon solute substitution, referenced to a corresponding substitution in a bulk (crystalline) site:

$$\Delta E_i^{\mathrm{seg}} = E_i^{\mathrm{GB}} - E^{\mathrm{Bulk}} \quad (1)$$

A negative segregation energy therefore represents a thermodynamic tendency for segregation. As seen in Fig. 2, APM shows excellent accuracy; in fact, the accuracy here is of a similar magnitude to that achieved by a multiscale method with a quantum-mechanical calculation at the core (QM/MM) [11]. Not shown in Fig. 2 is a similar set of validations we have performed on free surface sites, delivering a similar degree of confidence in the potential, over even less symmetric sites [72]. We thus conclude that APM can deliver quantum-like accuracy to resolve defect thermodynamics spanning a range over 100 kJ/mol.

As a first application of APM, we have produced GB segregation spectra spanning the elements in the periodic table when each is substituted into fcc Ni. Fig. 3 shows this full catalog of results as a dilute-limit segregation enthalpy ($\Delta E^{\text{seg}}$). The spectra are shown for nearly all possible solute species available with the universal interatomic potential. In each case we fit the segregation energy spectrum to a skew-normal [73,74] probability density distribution, $P(\Delta E^{\text{seg}})$, which has been shown to capture general segregation spectra very well in polycrystalline metals [10,23]:

$$P(\Delta E^{\text{seg}}) = \frac{1}{\sqrt{2\pi}\sigma} \exp\left[-\frac{(\Delta E^{\text{seg}} - \mu)^2}{2\sigma^2}\right] \text{erfc}\left[\frac{\alpha(\Delta E^{\text{seg}} - \mu)}{\sqrt{2}\sigma}\right] \tag{2}$$

where $\alpha$, $\mu$ and $\sigma$ are the skewness, location and width parameter of the distribution respectively. For all of the Ni alloys in Fig. 3, the set of fitted distribution parameters are collected in the supplemental material [48].

This is not the first time that a segregation spectrum such as shown in Fig. 2 has been reported; our prior work has done this using (i) only highly scalable potentials, with relatively low accuracy [10], and (ii) with multiscale methods using a quantum mechanical approach and very high computational cost [9]. The major advantage of the present APM approach is that it has far better accuracy and can address many more systems than method (i), and can proceed far beyond the computational reach of method (ii). This is addressed even more explicitly in Fig. 4, where we address the more computationally intensive challenge of segregation in bcc Fe. Although iron and steels are the predominant structural materials in the world [75], they are very difficult to model at the atomistic level due to the complexity of their electron structure and the influence of magnetism. The use of DFT multiscale methods on such problems is, at the time of this writing, not plausible in a manner that could produce a GB segregation spectrum [9,76]; APM thus provides the first database for segregation spectra in Fe with state-of-the-art accuracy.

Repeating these APM calculations for many systems, the spectra for 1,036 alloys (such as Fig. 1g) including the solvent of Ag [77], Al [78], Au [79], Cr [80], Cu [81], Fe [80], Mo [82], Nb [83], Ni [84], Pd [85], Pt [86], Ta [87], V [88] and W [89] are provided in the supplemental material [48]. For reference, this is roughly five times larger than the number of systems addressed previously with classic interatomic potentials [10], and roughly 25 times larger than the number of systems addressed with quantum-accurate methods [9]. The new database thus opens up a wide range of opportunities for developing advanced materials through interfacial engineering [86].

Another segregation problem that is presently far outside the range of computational plausibility for quantum-accurate multiscale models is the solute-solute interaction spectrum. At any given defect site, the local segregation energy is shifted if a nearby site is also occupied by solute. The interaction coefficient for a single site can be computed via [24,90]:

$$\omega_i^{\text{GB}} = \frac{1}{2}\sum_j \left[\left(E_{i-j}^{AA} + E_{i-j}^{BB}\right) - \left(E_{i-j}^{BA} + E_{i-j}^{AB}\right)\right] \tag{3}$$

where $\omega_i^{\text{GB}}$ is the interaction coefficient of site *i* computed for all nearest neighbor sites *j* using combinatoric pairwise energetics of solvent (A) and solute (B) at both sites (the system potential energies in Eq. (3) are denoted by whether site *i* and *j* are occupied by A or B respectively). This interaction parameter is a site-averaged pair-wise distribution for the first nearest neighbor [12,24], and while the method is, in principle, applicable to multi-body interactions as well as second and third nearest neighbor interactions, these extensions are left for future work. Even to compute the summation in Eq. (3) for all of the sites in a GB at the first nearest-neighbor level is a problem that is ~50-100 times larger than simply computing the segregation energy for a single site. As a result, the current state of the art on this particular thermodynamic term is entirely in mode (i) from above: only potentials have produced full interaction spectra [12,91]. The current APM method can address this problem with speed and chemical flexibility. And with the rapidly increasing accuracy of universal interatomic potentials, near-quantum accurate GB spectra are imminently accessible using APM.

We show a computed solute-solute interaction spectrum in Cu(Au) GBs from a high fidelity APM model (with 500 sampled sites); this computation requires ~200 times more computational resource than a single dilute segregation energy spectrum shown in Fig. 3 and 4 due to the coordination (Z~>12 at GB sites) and three configurations needed for each nearest neighbor pair [24]. The resulting distribution, seen in Fig. 5, can be described as a bivariate Gaussian:

$$P(\boldsymbol{x} = [\Delta E^{\mathrm{seg}}, \omega^{\mathrm{GB}}]) = \frac{1}{2\pi\sqrt{|\boldsymbol{\Sigma}|}} \exp\left[-\frac{1}{2}(\boldsymbol{x}-\boldsymbol{\mu})^T \boldsymbol{\Sigma}^{-1}(\boldsymbol{x}-\boldsymbol{\mu})\right] \tag{4}$$

with $\boldsymbol{\mu}$ and $\boldsymbol{\Sigma}$ as the mean and covariance matrix respectively for the random variable $\Delta E^{\mathrm{seg}}$ and $\omega^{\mathrm{GB}}$; the fitted parameters for the Cu(Au) system are given in Fig. 5. These kinds of interactions are often needed for accurate predictions of the segregated state in real materials when concentrations exceed the dilute limit: a brief estimation using the interaction of order 100 kJ/mol, as seen in Cu(Au) in Fig. 5, at a concentration $X^{\mathrm{GB}}$ of ~10 at. %, can give an energetic penalty of order 10 kJ/mol, which is close to the order of magnitude of the segregation energy spectrum itself. Spectra such as those of Figs. 3 and 4 thus need to be amplified to account for interactions, and the APM approach opens a computationally plausible pathway towards the construction of a full GB spectral genome for polycrystalline alloys [92].

In conclusion, we have developed and applied an augmented potential method (APM) for modeling defect chemistry by bridging multiscale interatomic potentials. Not only is the framework applicable to the construction of the GB spectral genome, but the method should also be applicable to other types of defects such as triple junctions [93–96], dislocations [97,98] and vacancy clusters [99,100]. With some modification, the method should also apply to interstitial elements [23,101] and their interactions, and should also pave the way towards more compositionally complex alloys [102–104]. We hope that the accelerated method will lead to the construction of a full defect genome [90,105,106] for materials science, enabling defect engineering in materials with large structural and chemical degrees of freedom.

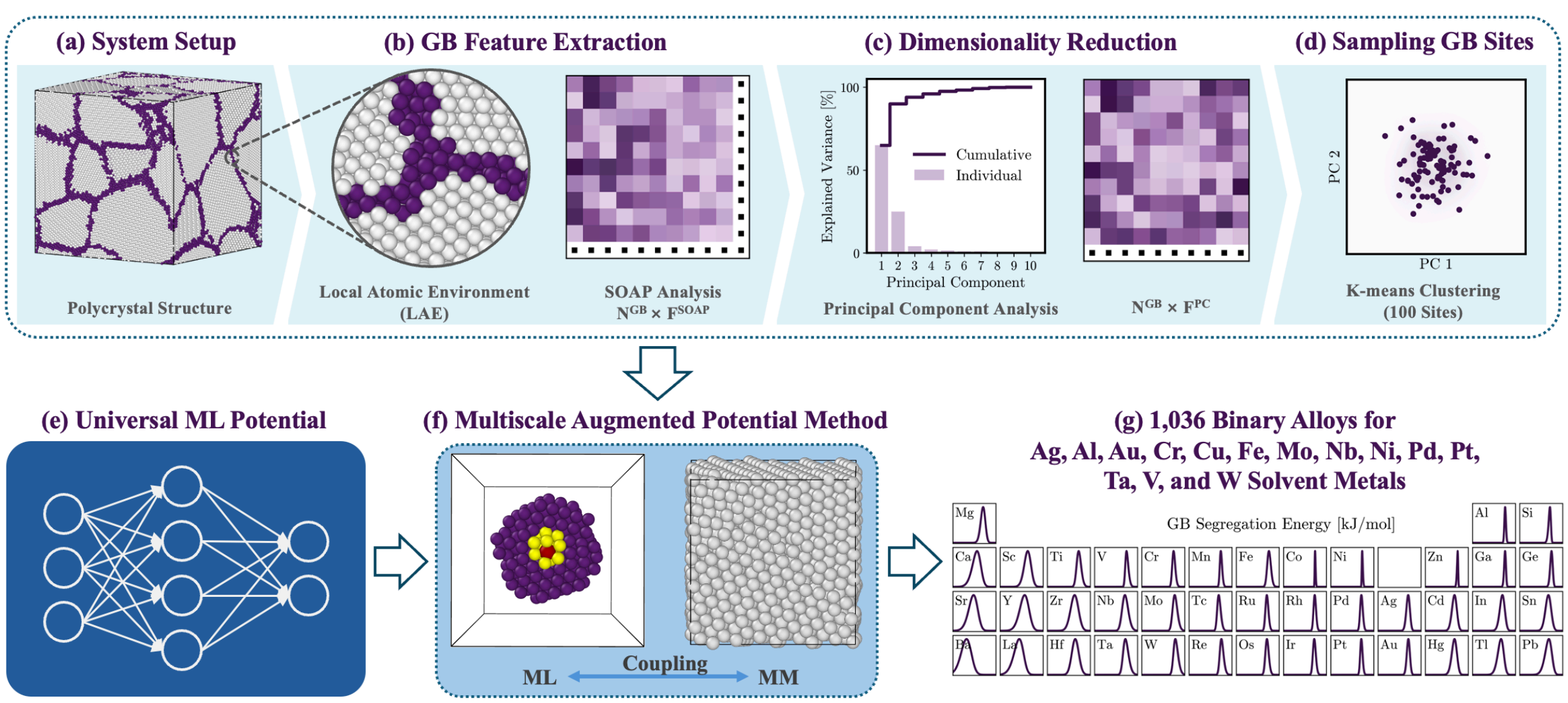

**Fig. 1.** A summary of the accelerated augmented potential method (APM) framework. The polycrystals generated in (a) with their features calculated in (b) are used to sample GB sites via dimensional reduction in (c) and (d). The APM method then takes a multiscale approach where the force field in the core part is replaced via a universal ML potential (e) to overcome the chemical boundary in traditional multiscale coupling methods that mix potentials (f). This allows construction of a defect spectral genome consisting of 1,036 grain boundary segregation spectra in (g). See Supplemental Material for details on the APM computation [48].

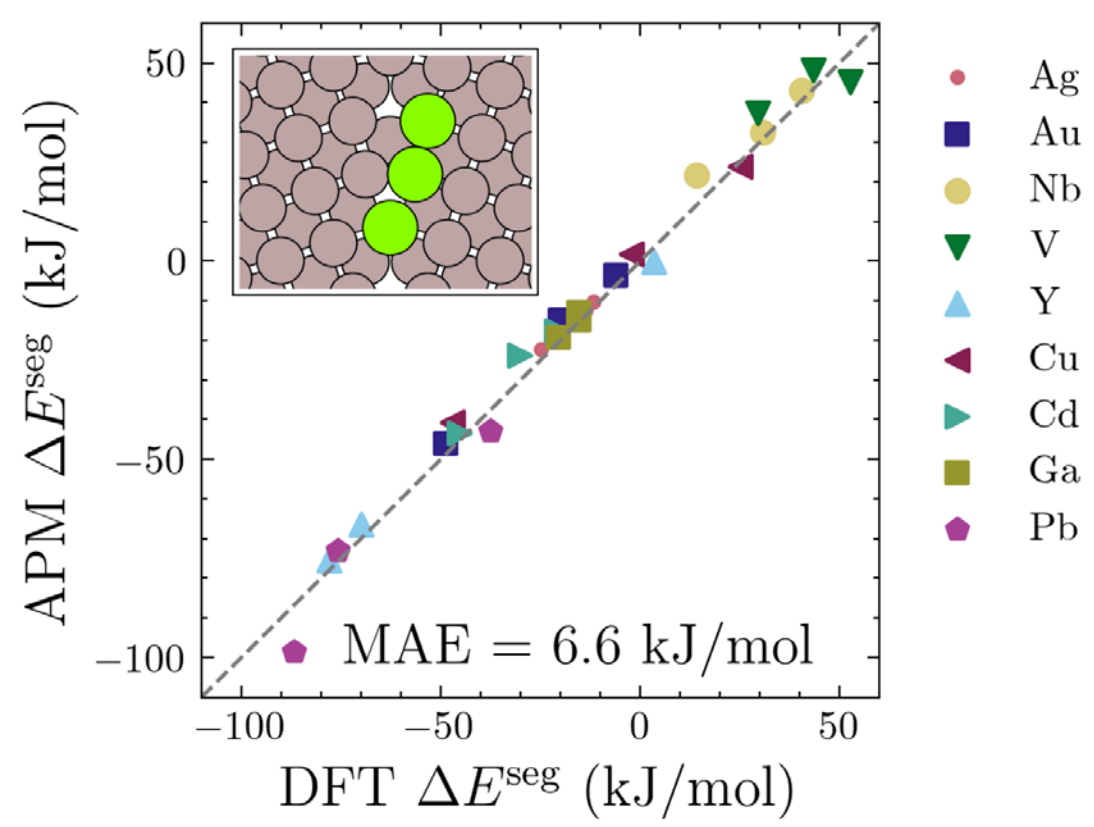

**Fig. 2.** The segregation energies computed from APM are then compared with DFT for Ag, Au, Nb, V, Y, Cu, Cd, Ga and Pb at Al Σ5[001](210) grain boundary using the three unique sites shown here. See Supplemental Material for details on the APM and DFT computation [48].

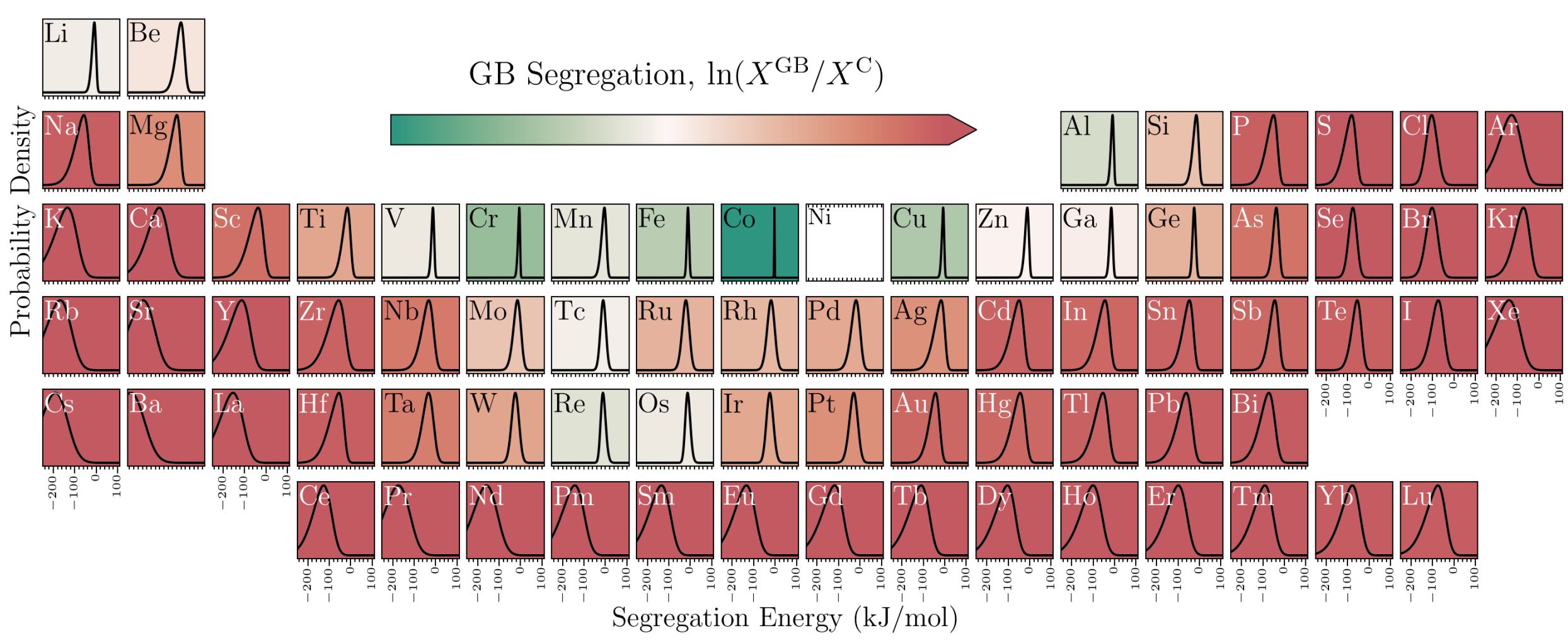

**Fig. 3.** Ni-based grain boundary spectra for 74 alloying elements. Details on the calculation of the segregation strengths and the distribution parameters can be found in the supplemental material [48].

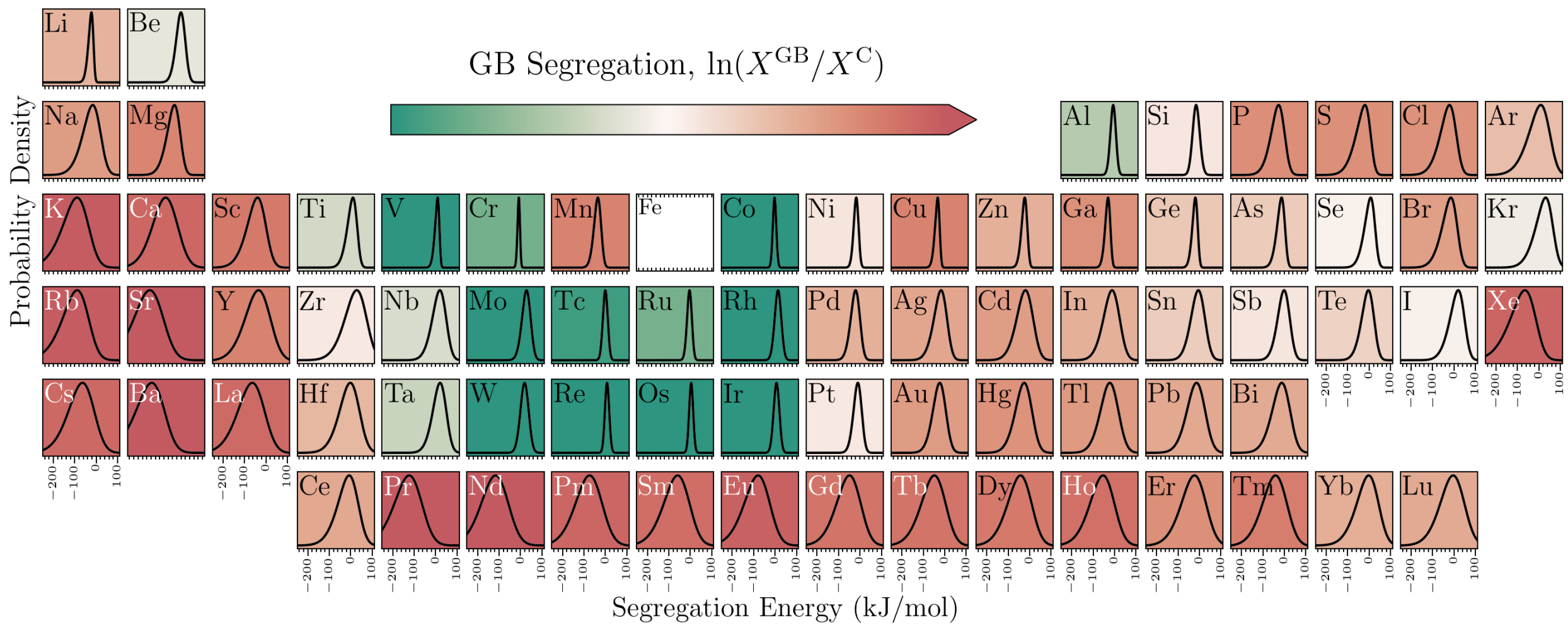

**Fig. 4.** Fe-based segregation spectra for 74 binary alloy pairs. Details on the calculation of the segregation strengths and the distribution parameters can be found in the supplemental material [48].

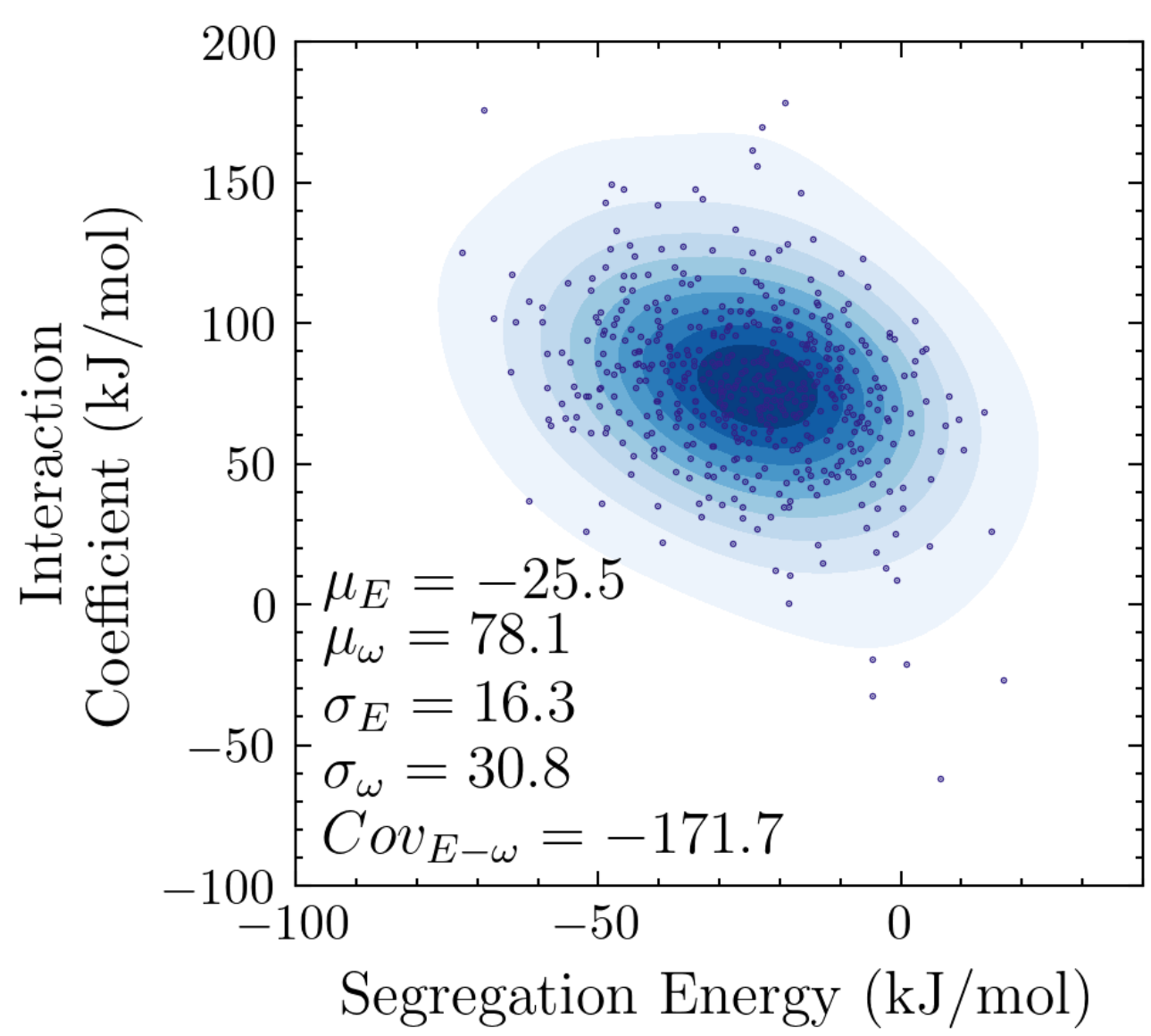

**Fig. 5.** The spectrum of solute (Au-Au) interaction energies in the GBs of Cu. The magnitude of interactions, which spans up to and beyond 100 kJ/mol, stresses the importance of solute-solute interactions in defect chemistry.

**CRediT authorship contribution statement**

**Nutth Tuchinda:** Conceptualization, Data curation, Formal analysis, Investigation, Methodology, Software, Validation, Visualization, Writing – original draft, Writing – review & editing. **Changle Li**: Formal Analysis, Investigation, Methodology, Software, Validation, Visualization, Writing – review & editing. **Christopher A. Schuh**: Conceptualization, Formal Analysis, Funding acquisition, Investigation,

Project administration, Resources, Supervision, Validation, Writing – original draft, Writing – review & editing

**Declaration of competing interest**

The authors declare that they have no known competing financial interests or personal relationships that could have appeared to influence the work reported in this paper.

**Acknowledgements**

This work was supported under the US Department of Energy award No. DE-SC0020180 (for the methodology development and grain boundary spectra) and Office of Naval Research (ONR) under the grant N000142312004 for the GB database of Fe-based systems. C. Li is supported by the Knut and Alice Wallenberg Foundation. The authors acknowledge MIT Satori and Research Computing Project for the computational resources used in this work. The authors also acknowledge the MIT SuperCloud and Lincoln Laboratory Supercomputing Center for providing HPC resources that have contributed to the research results reported within this paper [107].

**Data availability**

The data that support the findings of this article are available in the supplemental material [48].

**References**

[1] E.D. Hondros, M.P. Seah, The theory of grain boundary segregation in terms of surface adsorption analogues, MTA 8 (1977) 1363–1371. https://doi.org/10.1007/BF02642850.
[2] M. Alkayyali, F. Abdeljawad, Grain Boundary Solute Drag Model in Regular Solution Alloys, Phys. Rev. Lett. 127 (2021) 175503. https://doi.org/10.1103/PhysRevLett.127.175503.
[3] P. Lejček, Grain Boundary Segregation in Metals, Springer Berlin Heidelberg, Berlin, Heidelberg, 2010. https://doi.org/10.1007/978-3-642-12505-8.
[4] R.K. Koju, Y. Mishin, Relationship between grain boundary segregation and grain boundary diffusion in Cu-Ag alloys, Phys. Rev. Materials 4 (2020) 073403. https://doi.org/10.1103/PhysRevMaterials.4.073403.
[5] A. Argon, Strengthening mechanisms in crystal plasticity, OUP Oxford, 2007.
[6] T.J. Rupert, J.C. Trenkle, C.A. Schuh, Enhanced solid solution effects on the strength of nanocrystalline alloys, Acta Materialia 59 (2011) 1619–1631. https://doi.org/10.1016/j.actamat.2010.11.026.
[7] G.P.M. Leyson, W.A. Curtin, L.G. Hector, C.F. Woodward, Quantitative prediction of solute strengthening in aluminium alloys, Nature Mater 9 (2010) 750–755. https://doi.org/10.1038/nmat2813.
[8] M. Wagih, C.A. Schuh, Can symmetric tilt grain boundaries represent polycrystals?, Scripta Materialia 237 (2023) 115716.
[9] M. Wagih, C.A. Schuh, Learning Grain-Boundary Segregation: From First Principles to Polycrystals, Phys. Rev. Lett. 129 (2022) 046102. https://doi.org/10.1103/PhysRevLett.129.046102.
[10] M. Wagih, P.M. Larsen, C.A. Schuh, Learning grain boundary segregation energy spectra in polycrystals, Nat Commun 11 (2020) 6376. https://doi.org/10.1038/s41467-020-20083-6.
[11] N. Tuchinda, C.A. Schuh, Computed entropy spectra for grain boundary segregation in polycrystals, Npj Computational Materials 10 (2024) 72. https://doi.org/10.1038/s41524-024-01260-3.

[12] T.P. Matson, C.A. Schuh, A "bond-focused" local atomic environment representation for a high throughput solute interaction spectrum analysis, Acta Materialia 278 (2024) 120275. https://doi.org/10.1016/j.actamat.2024.120275.
[13] N. Bernstein, J.R. Kermode, G. Csányi, Hybrid atomistic simulation methods for materials systems, Reports on Progress in Physics 72 (2009) 026501. https://doi.org/10.1088/0034-4885/72/2/026501.
[14] L. Huber, B. Grabowski, M. Militzer, J. Neugebauer, J. Rottler, A QM/MM approach for low-symmetry defects in metals, Computational Materials Science 118 (2016) 259–268. https://doi.org/10.1016/j.commatsci.2016.03.028.
[15] N. Tuchinda, C.A. Schuh, The vibrational entropy spectra of grain boundary segregation in polycrystals, Acta Materialia 245 (2023) 118630. https://doi.org/10.1016/j.actamat.2022.118630.
[16] N. Choly, G. Lu, W. E, E. Kaxiras, Multiscale simulations in simple metals: A density-functional-based methodology, Phys. Rev. B 71 (2005) 094101. https://doi.org/10.1103/PhysRevB.71.094101.
[17] D. Gehringer, L. Huber, J. Neugebauer, D. Holec, Segregation to α 2 / γ interfaces in TiAl alloys: A multiscale QM/MM study, Phys. Rev. Materials 7 (2023) 063604. https://doi.org/10.1103/PhysRevMaterials.7.063604.
[18] Y. Liu, G. Lu, Z. Chen, N. Kioussis, An improved QM/MM approach for metals, Modelling and Simulation in Materials Science and Engineering 15 (2007) 275–284. https://doi.org/10.1088/0965-0393/15/3/006.
[19] P. Grigorev, A.M. Goryaeva, M.-C. Marinica, J.R. Kermode, T.D. Swinburne, Calculation of dislocation binding to helium-vacancy defects in tungsten using hybrid ab initio-machine learning methods, Acta Materialia 247 (2023) 118734. https://doi.org/10.1016/j.actamat.2023.118734.
[20] T.D. Swinburne, J.R. Kermode, Computing energy barriers for rare events from hybrid quantum/classical simulations through the virtual work principle, Phys. Rev. B 96 (2017) 144102. https://doi.org/10.1103/PhysRevB.96.144102.
[21] J.E. Saal, S. Kirklin, M. Aykol, B. Meredig, C. Wolverton, Materials Design and Discovery with High-Throughput Density Functional Theory: The Open Quantum Materials Database (OQMD), JOM 65 (2013) 1501–1509. https://doi.org/10.1007/s11837-013-0755-4.
[22] H. Zheng, X.-G. Li, R. Tran, C. Chen, M. Horton, D. Winston, K.A. Persson, S.P. Ong, Grain boundary properties of elemental metals, Acta Materialia 186 (2020) 40–49. https://doi.org/10.1016/j.actamat.2019.12.030.
[23] M. Wagih, C.A. Schuh, Spectrum of grain boundary segregation energies in a polycrystal, Acta Materialia 181 (2019) 228–237. https://doi.org/10.1016/j.actamat.2019.09.034.
[24] T.P. Matson, C.A. Schuh, Atomistic Assessment of Solute-Solute Interactions during Grain Boundary Segregation, Nanomaterials 11 (2021) 2360. https://doi.org/10.3390/nano11092360.
[25] K. Ito, H. Sawada, S. Ogata, Theoretical Prediction of Grain Boundary Segregation Using Nano-Polycrystalline Grain Boundary Model, Mater. Trans. 62 (2021) 575–581. https://doi.org/10.2320/matertrans.MT-M2020352.
[26] K. Ito, Y. Tanaka, H. Sawada, Application of Grain Boundary Segregation Prediction Using a Nano-Polycrystalline Grain Boundary Model to Transition Metal Solute Elements: Prediction of Grain Boundary Segregation of Mn and Cr in bcc-Fe Polycrystals, Mater. Trans. 63 (2022) 269–277. https://doi.org/10.2320/matertrans.MT-M2021205.
[27] K. Ito, Y. Tanaka, K. Tsutsui, H. Sawada, Analysis of grain-boundary segregation of hydrogen in bcc-Fe polycrystals via a nano-polycrystalline grain-boundary model, Computational Materials Science 225 (2023) 112196. https://doi.org/10.1016/j.commatsci.2023.112196.
[28] G. vanRossum, Python reference manual, Department of Computer Science [CS] (1995).
[29] A.H. Larsen, J.J. Mortensen, J. Blomqvist, I.E. Castelli, R. Christensen, M. Dułak, J. Friis, M.N. Groves, B. Hammer, C. Hargus, The atomic simulation environment—a Python library for working with atoms, Journal of Physics: Condensed Matter 29 (2017) 273002.

[30] F. Pedregosa, G. Varoquaux, A. Gramfort, V. Michel, B. Thirion, O. Grisel, M. Blondel, P. Prettenhofer, R. Weiss, V. Dubourg, Scikit-learn: Machine learning in Python, The Journal of Machine Learning Research 12 (2011) 2825–2830.
[31] C.R. Harris, K.J. Millman, S.J. van der Walt, R. Gommers, P. Virtanen, D. Cournapeau, E. Wieser, J. Taylor, S. Berg, N.J. Smith, R. Kern, M. Picus, S. Hoyer, M.H. van Kerkwijk, M. Brett, A. Haldane, J.F. del Río, M. Wiebe, P. Peterson, P. Gérard-Marchant, K. Sheppard, T. Reddy, W. Weckesser, H. Abbasi, C. Gohlke, T.E. Oliphant, Array programming with NumPy, Nature 585 (2020) 357–362. https://doi.org/10.1038/s41586-020-2649-2.
[32] P. Virtanen, R. Gommers, T.E. Oliphant, M. Haberland, T. Reddy, D. Cournapeau, E. Burovski, P. Peterson, W. Weckesser, J. Bright, S.J. van der Walt, M. Brett, J. Wilson, K.J. Millman, N. Mayorov, A.R.J. Nelson, E. Jones, R. Kern, E. Larson, C.J. Carey, İ. Polat, Y. Feng, E.W. Moore, J. VanderPlas, D. Laxalde, J. Perktold, R. Cimrman, I. Henriksen, E.A. Quintero, C.R. Harris, A.M. Archibald, A.H. Ribeiro, F. Pedregosa, P. van Mulbregt, A. Vijaykumar, A.P. Bardelli, A. Rothberg, A. Hilboll, A. Kloeckner, A. Scopatz, A. Lee, A. Rokem, C.N. Woods, C. Fulton, C. Masson, C. Häggström, C. Fitzgerald, D.A. Nicholson, D.R. Hagen, D.V. Pasechnik, E. Olivetti, E. Martin, E. Wieser, F. Silva, F. Lenders, F. Wilhelm, G. Young, G.A. Price, G.-L. Ingold, G.E. Allen, G.R. Lee, H. Audren, I. Probst, J.P. Dietrich, J. Silterra, J.T. Webber, J. Slavič, J. Nothman, J. Buchner, J. Kulick, J.L. Schönberger, J.V. de Miranda Cardoso, J. Reimer, J. Harrington, J.L.C. Rodríguez, J. Nunez-Iglesias, J. Kuczynski, K. Tritz, M. Thoma, M. Newville, M. Kümmerer, M. Bolingbroke, M. Tartre, M. Pak, N.J. Smith, N. Nowaczyk, N. Shebanov, O. Pavlyk, P.A. Brodtkorb, P. Lee, R.T. McGibbon, R. Feldbauer, S. Lewis, S. Tygier, S. Sievert, S. Vigna, S. Peterson, S. More, T. Pudlik, T. Oshima, T.J. Pingel, T.P. Robitaille, T. Spura, T.R. Jones, T. Cera, T. Leslie, T. Zito, T. Krauss, U. Upadhyay, Y.O. Halchenko, Y. Vázquez-Baeza, SciPy 1.0 Contributors, SciPy 1.0: fundamental algorithms for scientific computing in Python, Nature Methods 17 (2020) 261–272. https://doi.org/10.1038/s41592-019-0686-2.
[33] J.D. Hunter, Matplotlib: A 2D graphics environment, Computing in Science & Engineering 9 (2007) 90–95. https://doi.org/10.1109/MCSE.2007.55.
[34] John Garrett, Echedey Luis, H. -H. Peng, Tim Cera, gobinathj, Josh Borrow, Mehmet Keçeci, Splines, Suraj Iyer, Yuming Liu, cjw, Mikhail Gasanov, garrettj403/SciencePlots: 2.1.1, (2023). https://doi.org/10.5281/ZENODO.10206719.
[35] K.M. Thyng, C.A. Greene, R.D. Hetland, H.M. Zimmerle, S.F. DiMarco, True Colors of Oceanography, Oceanography 29 (2016) 9–13.
[36] M.L. Waskom, Seaborn: statistical data visualization, Journal of Open Source Software 6 (2021) 3021.
[37] A. Stukowski, Visualization and analysis of atomistic simulation data with OVITO–the Open Visualization Tool, Modelling and Simulation in Materials Science and Engineering 18 (2009) 015012. https://doi.org/10.1088/0965-0393/18/1/015012.
[38] R. Hadian, B. Grabowski, J. Neugebauer, GB code: A grain boundary generation code, The Journal of Open Source Software 3 (2018). https://doi.org/10.21105/joss.00900.
[39] T. Kluyver, B. Ragan-Kelley, F. Pérez, B. Granger, M. Bussonnier, J. Frederic, K. Kelley, J. Hamrick, J. Grout, S. Corlay, P. Ivanov, D. Avila, S. Abdalla, C. Willing, Jupyter Notebooks – a publishing format for reproducible computational workflows, in: F. Loizides, B. Schmidt (Eds.), Positioning and Power in Academic Publishing: Players, Agents and Agendas, IOS Press, 2016: pp. 87–90.
[40] A.P. Bartók, R. Kondor, G. Csányi, On representing chemical environments, Phys. Rev. B 87 (2013) 184115. https://doi.org/10.1103/PhysRevB.87.184115.
[41] J.R. Kermode, f90wrap: an automated tool for constructing deep Python interfaces to modern Fortran codes, Journal of Physics: Condensed Matter 32 (2020) 305901.
[42] G. Csányi, S. Winfield, J. Kermode, M. Payne, A. Comisso, A. De Vita, N. Bernstein, Expressive programming for computational physics in Fortran 950+, Newsletter of the Computational Physics Group (2007) 1–24.

[43] H. Hotelling, Analysis of a complex of statistical variables into principal components., Journal of Educational Psychology 24 (1933) 417.
[44] I.T. Jolliffe, A note on the use of principal components in regression, Journal of the Royal Statistical Society Series C: Applied Statistics 31 (1982) 300–303.
[45] M.E. Tipping, C.M. Bishop, Probabilistic principal component analysis, Journal of the Royal Statistical Society: Series B (Statistical Methodology) 61 (1999) 611–622.
[46] S. Lloyd, Least squares quantization in PCM, IEEE Transactions on Information Theory 28 (1982) 129–137.
[47] C. Elkan, Using the triangle inequality to accelerate k-means, in: Proceedings of the 20th International Conference on Machine Learning (ICML-03), 2003: pp. 147–153.
[48] See Supplemental Material at [URL will be inserted by publisher] for the details of validations, calculations and the supplementary software. (n.d.).
[49] Y.-L. Liao, B. Wood, A. Das, T. Smidt, Equiformerv2: Improved equivariant transformer for scaling to higher-degree representations, arXiv Preprint arXiv:2306.12059 (2023).
[50] J. Riebesell, R.E. Goodall, A. Jain, P. Benner, K.A. Persson, A.A. Lee, Matbench Discovery--An evaluation framework for machine learning crystal stability prediction, arXiv Preprint arXiv:2308.14920 (2023).
[51] A. Dunn, Q. Wang, A. Ganose, D. Dopp, A. Jain, Benchmarking materials property prediction methods: the Matbench test set and Automatminer reference algorithm, Npj Comput Mater 6 (2020) 138. https://doi.org/10.1038/s41524-020-00406-3.
[52] L. Barroso-Luque, M. Shuaibi, X. Fu, B.M. Wood, M. Dzamba, M. Gao, A. Rizvi, C.L. Zitnick, Z.W. Ulissi, Open materials 2024 (omat24) inorganic materials dataset and models, arXiv Preprint arXiv:2410.12771 (2024).
[53] B. Focassio, L.P. M. Freitas, G.R. Schleder, Performance assessment of universal machine learning interatomic potentials: Challenges and directions for materials' surfaces, ACS Applied Materials & Interfaces (2024).
[54] I. Batatia, D.P. Kovacs, G. Simm, C. Ortner, G. Csányi, MACE: Higher order equivariant message passing neural networks for fast and accurate force fields, Advances in Neural Information Processing Systems 35 (2022) 11423–11436.
[55] C.A. Becker, F. Tavazza, Z.T. Trautt, R.A. Buarque de Macedo, Considerations for choosing and using force fields and interatomic potentials in materials science and engineering, Current Opinion in Solid State and Materials Science 17 (2013) 277–283. https://doi.org/10.1016/j.cossms.2013.10.001.
[56] S. Batzner, A. Musaelian, L. Sun, M. Geiger, J.P. Mailoa, M. Kornbluth, N. Molinari, T.E. Smidt, B. Kozinsky, E (3)-equivariant graph neural networks for data-efficient and accurate interatomic potentials, Nature Communications 13 (2022) 2453.
[57] I. Batatia, P. Benner, Y. Chiang, A.M. Elena, D.P. Kovács, J. Riebesell, X.R. Advincula, M. Asta, M. Avaylon, W.J. Baldwin, A foundation model for atomistic materials chemistry, arXiv Preprint arXiv:2401.00096 (2023).
[58] Y.-L. Liao, T. Smidt, Equiformer: Equivariant graph attention transformer for 3d atomistic graphs, arXiv Preprint arXiv:2206.11990 (2022).
[59] S. Passaro, C.L. Zitnick, Reducing SO (3) convolutions to SO (2) for efficient equivariant GNNs, in: International Conference on Machine Learning, PMLR, 2023: pp. 27420–27438.
[60] J. Schmidt, N. Hoffmann, H.-C. Wang, P. Borlido, P.J. Carriço, T.F. Cerqueira, S. Botti, M.A. Marques, Machine-Learning-Assisted Determination of the Global Zero-Temperature Phase Diagram of Materials, Advanced Materials 35 (2023) 2210788.
[61] A. Jain, S.P. Ong, G. Hautier, W. Chen, W.D. Richards, S. Dacek, S. Cholia, D. Gunter, D. Skinner, G. Ceder, K.A. Persson, Commentary: The Materials Project: A materials genome approach to accelerating materials innovation, APL Materials 1 (2013) 011002. https://doi.org/10.1063/1.4812323.

[62] J. Schmidt, T.F.T. Cerqueira, A.H. Romero, A. Loew, F. Jäger, H.-C. Wang, S. Botti, M.A.L. Marques, Improving machine-learning models in materials science through large datasets, Materials Today Physics 48 (2024) 101560. https://doi.org/10.1016/j.mtphys.2024.101560.
[63] L. Chanussot, A. Das, S. Goyal, T. Lavril, M. Shuaibi, M. Riviere, K. Tran, J. Heras-Domingo, C. Ho, W. Hu, Open catalyst 2020 (OC20) dataset and community challenges, Acs Catalysis 11 (2021) 6059–6072.
[64] Yuan Chiang, mace-universal, (n.d.). https://doi.org/10.57967/HF/1202.
[65] I. Batatia, S. Batzner, D.P. Kovács, A. Musaelian, G.N.C. Simm, R. Drautz, C. Ortner, B. Kozinsky, G. Csányi, The design space of E(3)-equivariant atom-centred interatomic potentials, Nature Machine Intelligence 7 (2025) 56–67. https://doi.org/10.1038/s42256-024-00956-x.
[66] B. Deng, P. Zhong, K. Jun, J. Riebesell, K. Han, C.J. Bartel, G. Ceder, CHGNet: Pretrained universal neural network potential for charge-informed atomistic modeling, (2023). https://doi.org/10.48550/arXiv.2302.14231.
[67] A.P. Thompson, H.M. Aktulga, R. Berger, D.S. Bolintineanu, W.M. Brown, P.S. Crozier, P.J. in 't Veld, A. Kohlmeyer, S.G. Moore, T.D. Nguyen, R. Shan, M.J. Stevens, J. Tranchida, C. Trott, S.J. Plimpton, LAMMPS - a flexible simulation tool for particle-based materials modeling at the atomic, meso, and continuum scales, Computer Physics Communications 271 (2022) 108171. https://doi.org/10.1016/j.cpc.2021.108171.
[68] W.M. Brown, A. Kohlmeyer, S.J. Plimpton, A.N. Tharrington, Implementing molecular dynamics on hybrid high performance computers–Particle–particle particle-mesh, Computer Physics Communications 183 (2012) 449–459. https://doi.org/10.1016/j.cpc.2011.10.012.
[69] W.M. Brown, P. Wang, S.J. Plimpton, A.N. Tharrington, Implementing molecular dynamics on hybrid high performance computers–short range forces, Computer Physics Communications 182 (2011) 898–911. https://doi.org/10.1016/j.cpc.2010.12.021.
[70] W.M. Brown, M. Yamada, Implementing molecular dynamics on hybrid high performance computers—three-body potentials, Computer Physics Communications 184 (2013) 2785–2793. https://doi.org/10.1016/j.cpc.2013.08.002.
[71] N. Tuchinda, G.B. Olson, C.A. Schuh, Grain Boundary Segregation and Embrittlement of Aluminum Binary Alloys from First Principles, Acta Materialia (2025) 121058. https://doi.org/10.1016/j.actamat.2025.121058.
[72] N. Tuchinda, G.B. Olson, C.A. Schuh, A Grain Boundary Embrittlement Genome for Substitutional Cubic Alloys, (2025). https://doi.org/10.48550/arXiv.2502.06531.
[73] A. Azzalini, The R package "sn": The skew-normal and related distributions such as the skew-t and the SUN (version 2.0.2)., Università degli Studi di Padova, Italia, 2022. https://cran.r-project.org/package=sn.
[74] A. Azzalini, A. Capitanio, Statistical applications of the multivariate skew normal distribution, Journal of the Royal Statistical Society: Series B (Statistical Methodology) 61 (1999) 579–602. https://doi.org/10.1111/1467-9868.00194.
[75] H. Bhadeshia, R. Honeycombe, Steels: microstructure and properties, Butterworth-Heinemann, 2017.
[76] R. Freitas, Y. Cao, Machine-learning potentials for crystal defects, MRS Communications (2022) 1–11.
[77] P.L. Williams, Y. Mishin, J.C. Hamilton, An embedded-atom potential for the Cu–Ag system, Modelling and Simulation in Materials Science and Engineering 14 (2006) 817. https://doi.org/10.1088/0965-0393/14/5/002.
[78] M.I. Mendelev, M.J. Kramer, C.A. Becker, M. Asta, Analysis of semi-empirical interatomic potentials appropriate for simulation of crystalline and liquid Al and Cu, Philosophical Magazine 88 (2008) 1723–1750. https://doi.org/10.1080/14786430802206482.
[79] P.A.T. Olsson, Transverse resonant properties of strained gold nanowires, Journal of Applied Physics 108 (2010) 034318. https://doi.org/10.1063/1.3460127.

[80] S.M. Eich, D. Beinke, G. Schmitz, Embedded-atom potential for an accurate thermodynamic description of the iron–chromium system, Computational Materials Science 104 (2015) 185–192. https://doi.org/10.1016/j.commatsci.2015.03.047.
[81] M.I. Mendelev, A.H. King, The interactions of self-interstitials with twin boundaries, Philosophical Magazine 93 (2013) 1268–1278. https://doi.org/10.1080/14786435.2012.747012.
[82] Y. Chen, X. Liao, N. Gao, W. Hu, F. Gao, H. Deng, Interatomic potentials of W–V and W–Mo binary systems for point defects studies, Journal of Nuclear Materials 531 (2020) 152020. https://doi.org/10.1016/j.jnucmat.2020.152020.
[83] M.R. Fellinger, H. Park, J.W. Wilkins, Force-matched embedded-atom method potential for niobium, Phys. Rev. B 81 (2010) 144119. https://doi.org/10.1103/PhysRevB.81.144119.
[84] M.I. Mendelev, M.J. Kramer, S.G. Hao, K.M. Ho, C.Z. Wang, Development of interatomic potentials appropriate for simulation of liquid and glass properties of NiZr2 alloy, Philosophical Magazine 92 (2012) 4454–4469. https://doi.org/10.1080/14786435.2012.712220.
[85] S.M. Foiles, M.I. Baskes, M.S. Daw, Embedded-atom-method functions for the fcc metals Cu, Ag, Au, Ni, Pd, Pt, and their alloys, Phys. Rev. B 33 (1986) 7983–7991. https://doi.org/10.1103/PhysRevB.33.7983.
[86] A. Hirata, L.J. Kang, T. Fujita, B. Klumov, K. Matsue, M. Kotani, A.R. Yavari, M.W. Chen, Geometric frustration of icosahedron in metallic glasses, Science 341 (2013) 376–379.
[87] Y. Chen, J. Fang, L. Liu, W. Hu, N. Gao, F. Gao, H. Deng, Development of the interatomic potentials for W-Ta system, Computational Materials Science 163 (2019) 91–99. https://doi.org/10.1016/j.commatsci.2019.03.021.
[88] M.I. Mendelev, S. Han, W. Son, G.J. Ackland, D.J. Srolovitz, Simulation of the interaction between Fe impurities and point defects in V, Phys. Rev. B 76 (2007) 214105. https://doi.org/10.1103/PhysRevB.76.214105.
[89] D.R. Mason, D. Nguyen-Manh, C.S. Becquart, An empirical potential for simulating vacancy clusters in tungsten, Journal of Physics: Condensed Matter 29 (2017) 505501.
[90] T.P. Matson, C.A. Schuh, Phase and defect diagrams based on spectral grain boundary segregation: A regular solution approach, Acta Materialia 265 (2024) 119584. https://doi.org/10.1016/j.actamat.2023.119584.
[91] N. Tuchinda, C.A. Schuh, Grain size dependencies of intergranular solute segregation in nanocrystalline materials, Acta Materialia 226 (2022) 117614. https://doi.org/10.1016/j.actamat.2021.117614.
[92] Q. Yan, S. Kar, S. Chowdhury, A. Bansil, The Case for a Defect Genome Initiative, Advanced Materials 36 (2024) 2303098. https://doi.org/10.1002/adma.202303098.
[93] I. Adlakha, K.N. Solanki, Atomic-scale investigation of triple junction role on defects binding energetics and structural stability in α-Fe, Acta Materialia 118 (2016) 64–76. https://doi.org/10.1016/j.actamat.2016.07.026.
[94] S.M. Eich, G. Schmitz, Embedded-atom study of grain boundary segregation and grain boundary free energy in nanosized iron–chromium tricrystals, Acta Materialia 147 (2018) 350–364. https://doi.org/10.1016/j.actamat.2018.01.006.
[95] N. Tuchinda, C.A. Schuh, Triple Junction Excess Energy in Polycrystalline Metals, Acta Materialia (2024) 120274. https://doi.org/10.1016/j.actamat.2024.120274.
[96] A.K. Barnett, O. Hussein, M. Alghalayini, A. Hinojos, J.E.I. Nathaniel, D.L. Medlin, K. Hattar, B.L. Boyce, F. Abdeljawad, Triple Junction Segregation Dominates the Stability of Nanocrystalline Alloys, Nano Lett. 24 (2024) 9627–9634. https://doi.org/10.1021/acs.nanolett.4c02395.
[97] L. Ventelon, B. Lüthi, E. Clouet, L. Proville, B. Legrand, D. Rodney, F. Willaime, Dislocation core reconstruction induced by carbon segregation in bcc iron, Phys. Rev. B 91 (2015) 220102. https://doi.org/10.1103/PhysRevB.91.220102.

[98] R. Kirchheim, Reducing grain boundary, dislocation line and vacancy formation energies by solute segregation. I. Theoretical background, Acta Materialia 55 (2007) 5129–5138. https://doi.org/10.1016/j.actamat.2007.05.047.
[99] J. Peng, S. Bahl, A. Shyam, J.A. Haynes, D. Shin, Solute-vacancy clustering in aluminum, Acta Materialia 196 (2020) 747–758. https://doi.org/10.1016/j.actamat.2020.06.062.
[100] P. Yi, T.T. Sasaki, S. Eswarappa Prameela, T.P. Weihs, M.L. Falk, The interplay between solute atoms and vacancy clusters in magnesium alloys, Acta Materialia 249 (2023) 118805. https://doi.org/10.1016/j.actamat.2023.118805.
[101] Z. Zhang, C. Deng, Grain boundary interstitial segregation in substitutional binary alloys, Acta Materialia 291 (2025) 121019. https://doi.org/10.1016/j.actamat.2025.121019.
[102] D. Aksoy, J. Luo, P. Cao, T.J. Rupert, A machine learning framework for the prediction of grain boundary segregation in chemically complex environments, Modelling Simul. Mater. Sci. Eng. 32 (2024) 065011. https://doi.org/10.1088/1361-651X/ad585f.
[103] J. Luo, Grain boundary segregation models for high-entropy alloys: Theoretical formulation and application to elucidate high-entropy grain boundaries, Journal of Applied Physics 135 (2024) 165303. https://doi.org/10.1063/5.0200669.
[104] I. Geiger, Y. Tian, Y. Han, Y. Bi, X. Pan, P. Cao, T.J. Rupert, Grain boundaries amplify local chemical ordering in CrCoNi, (2025). https://doi.org/10.48550/arXiv.2501.03901.
[105] A. Tehranchi, S. Zhang, A. Zendegani, C. Scheu, T. Hickel, J. Neugebauer, Metastable defect phase diagrams as roadmap to tailor chemically driven defect formation, Acta Materialia 277 (2024) 120145. https://doi.org/10.1016/j.actamat.2024.120145.
[106] C. Hu, D.L. Medlin, R. Dingreville, Stability and Mobility of Disconnections in Solute Atmospheres: Insights from Interfacial Defect Diagrams, Phys. Rev. Lett. 134 (2025) 016202. https://doi.org/10.1103/PhysRevLett.134.016202.
[107] A. Reuther, J. Kepner, C. Byun, S. Samsi, W. Arcand, D. Bestor, B. Bergeron, V. Gadepally, M. Houle, M. Hubbell, M. Jones, A. Klein, L. Milechin, J. Mullen, A. Prout, A. Rosa, C. Yee, P. Michaleas, Interactive Supercomputing on 40,000 Cores for Machine Learning and Data Analysis, in: 2018 IEEE High Performance Extreme Computing Conference (HPEC), 2018: pp. 1–6. https://doi.org/10.1109/HPEC.2018.8547629.

# The Augmented Potential Method: Multiscale Modeling Toward a Spectral Defect Genome

Nutth Tuchinda[1], Changle Li[1], Christopher A. Schuh[1,2*]

[1]*Department of Materials Science and Engineering, Massachusetts Institute of Technology, 77 Massachusetts Avenue, Cambridge, MA, 02139, USA*

[2] *Department of Materials Science and Engineering, Northwestern University, 2145 Sheridan Road, Evanston, IL 60208, USA*

*Correspondence to: schuh@northwestern.edu

## Supplemental Material

### SM 1. Solvent Polycrystalline Models

The solvent polycrystals are generated via the atomsk package using a random seed with 12 grain and 150 Å on each side. The systems are then annealed for 500 ps and cooled at 3 K/ps (see table S1 for the annealing temperature and interatomic potential used). The timestep is kept at 1 fs. The molecular dynamics simulations are all conducted in LAMMPS [1–4].

TABLE S1. List of the metallic solvents, annealing temperature, core size, buffer size and the potentials used in this work.

| Solvent | Annealing Temperature (K) | Core size (Å) | Butter Size (Å) | Core Potential | MM Potential |
|---|---|---|---|---|---|
| Ag | 494 | 5.03 | 7.5 | eqV2_31M_omat_mp_salex[1] | [5] |
| Al | 373 | 4.87 | 7.3 | eqV2_31M_omat_mp_salex[1] | [6] |
| Au | 535 | 5.04 | 7.0 | eqV2_31M_omat_mp_salex[1] | [7] |
| Cr | 872 | 4.39 | 6.3 | eqV2_31M_omat_mp_salex[1] | [8] |
| Cu | 700 | 4.35 | 6.5 | eqV2_31M_omat_mp_salex[1] | [9] |
| Fe | 724 | 4.35 | 6.3 | eqV2_31M_omat_mp_salex[1] | [8] |
| Mo | 1158 | 4.86 | 9.7 | eqV2_31M_omat_mp_salex[1] | [10] |
| Nb | 1100 | 4.35 | 9.0 | eqV2_31M_omat_mp_salex[1] | [11] |
| Ni | 800 | 4.26 | 6.5 | eqV2_31M_omat_mp_salex[1] | [12] |
| Pd | 731 | 4.77 | 6.6 | eqV2_31M_omat_mp_salex[1] | [13] |
| Pt | 816 | 4.80 | 8.7 | eqV2_31M_omat_mp_salex[1] | [14] |
| Ta | 1317 | 5.14 | 11.2 | mace-omat-0-medium[2] | [15] |
| V | 873 | 4.6 | 9.7 | mace-omat-0-medium[2] | [16] |
| W | 1478 | 4.9 | 8.7 | eqV2_31M_omat_mp_salex[1] | [17] |

[1] eqV2_31M_omat_mp_salex [18–23]

[2] mace-omat [19,24–27]

### SM 2. Augmented Potential Methods

We apply a universal machine learning potential for the core and buffer calculations of potential energy and force (see Table S1 for the potentials used). We apply the FIRE minimization algorithm [28,29] with a force norm of 0.02 eV/Å (except 0.03 eV/Å for V), which are lower than are typically used for QMMM calculations in Ref. [30]. The MM potentials are also scaled according to Refs. [30,31] to reduce mechanical mismatch between the potentials. We plot the buffer size convergence for the forces in Fig. S1 for all solvent species listed in Table S1, showing a good convergence vis-à-vis QM calculations in Ref. [30]. The segregation energy is then calculated using:

$$\Delta E_i^{\mathrm{seg}} = (E_i^{\mathrm{GB,solute}} - E_i^{\mathrm{GB,solvent}}) - (E^{\mathrm{Bulk,solute}} - E^{\mathrm{Bulk,solvent}}) \tag{S1}$$

where the solute and solvent denote the APM calculation at a site type '$i$' or a bulk site with an identical set of core and buffer zone atoms around a given GB or bulk site.

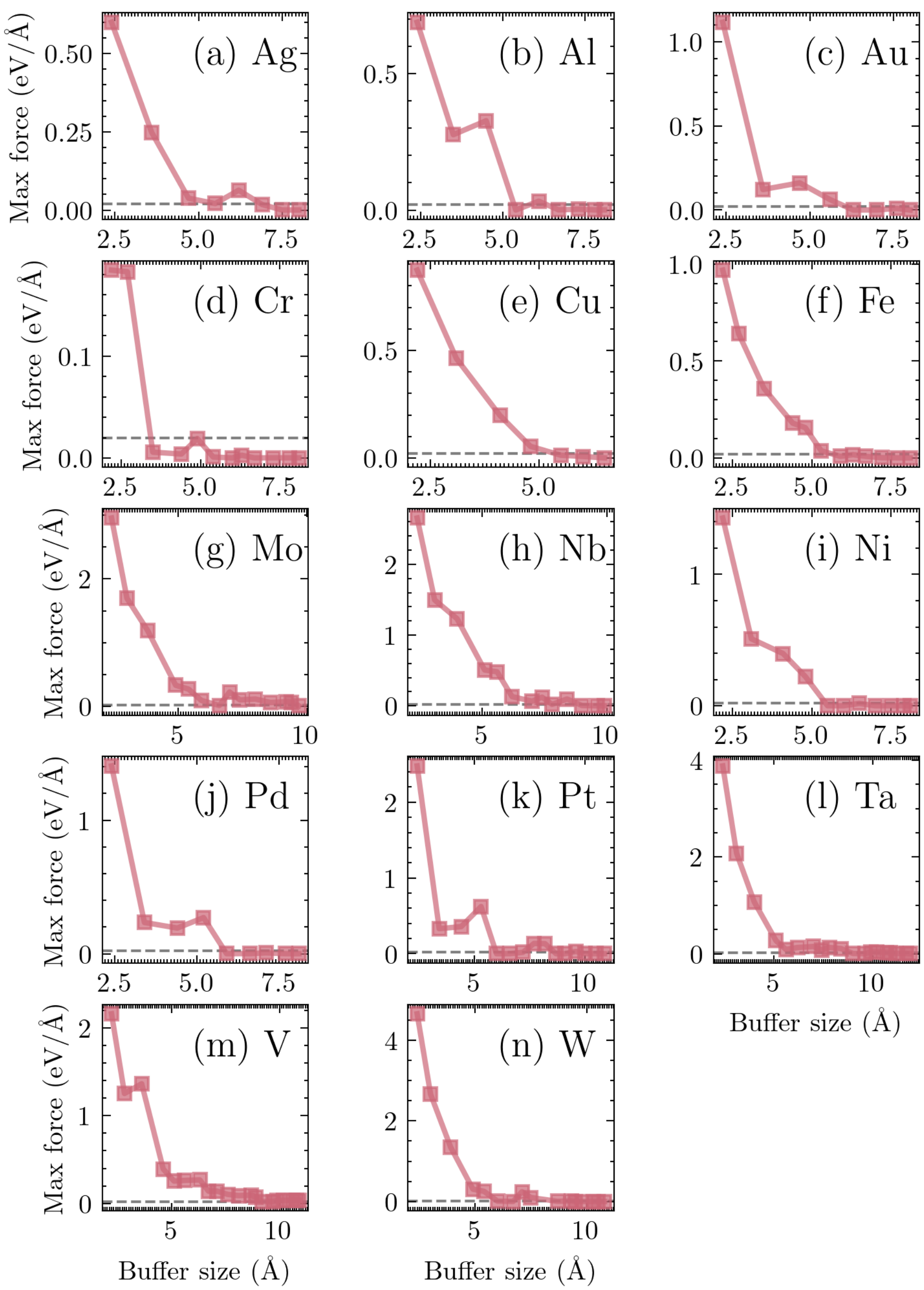

Fig. S1 Force matching as a function of buffer region size for all base element lattices.

### SM 3. Validation Test for Al Σ5[001](210)

We use the data reported in Ref. [32]. In essence, a 25-layer Σ5[001](210) GB slab is generated following Zhang *et al*. [33] using the gb_code [34] package, resulting in a GB of size 4.04×9.03 Å. The density functional theory computations are conducted via the Vienna Ab initio Software Package (VASP) [35–39] with k-point grid of 7×7, a planewave cutoff energy of 800 eV and the Perdew-Burke-Ernzerhof generalized gradient approximation functional [40]. The systems are relaxed to 0.01 eV/Å for the solute energies computed here.

### SM 4. Segregation Strength from Spectral Isotherm

The color scale of the segregation strength shown here is calculated as the natural logarithm of the segregation factor ($X^{\mathrm{GB}}/X^{\mathrm{C}}$) where the grain boundary concentration can be integrated from the spectra:

$$X^{\mathrm{GB}} = \int P\left[1 + \frac{1 - X^{\mathrm{C}}}{X^{\mathrm{C}}}\exp\left(\frac{\Delta E^{\mathrm{seg}}}{k_{\mathrm{B}}T}\right)\right]^{-1} d\Delta E^{\mathrm{seg}} \tag{S2}$$

with $k_{\mathrm{B}}$ and $T$ as the Boltzmann constant and temperature respectively. $P$ is the distribution of Eq. (2) in the main text. The temperatures in Fig. 3, 4 and S2-S17 are kept at half of the melting point (0.5 $T^{\mathrm{melt}}$).

### SM 5. Spectral Grain Boundary Genome

The grain boundary spectra are plotted here below, and their distribution parameters are also tabulated in a separate spreadsheet in the supplemental materials.

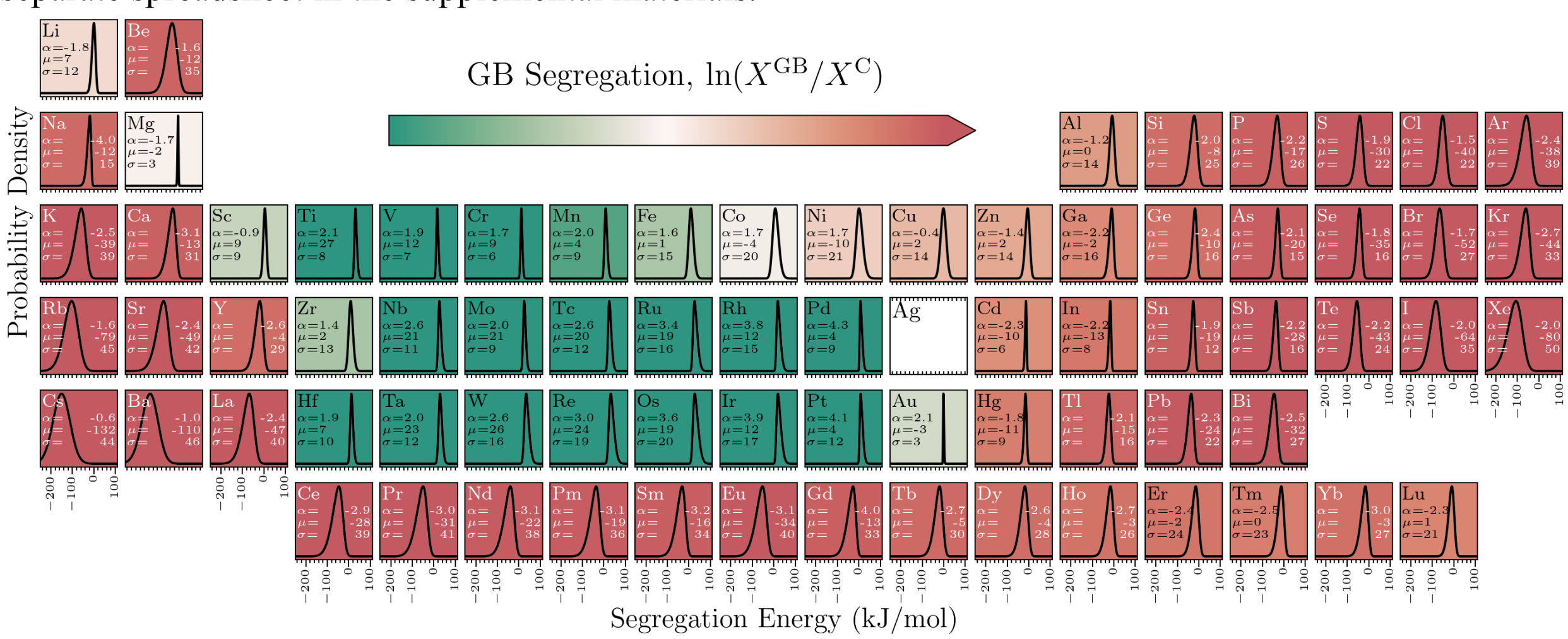

Fig. S2 Segregation spectra for Ag-based substitutional binary alloys in kJ/mol

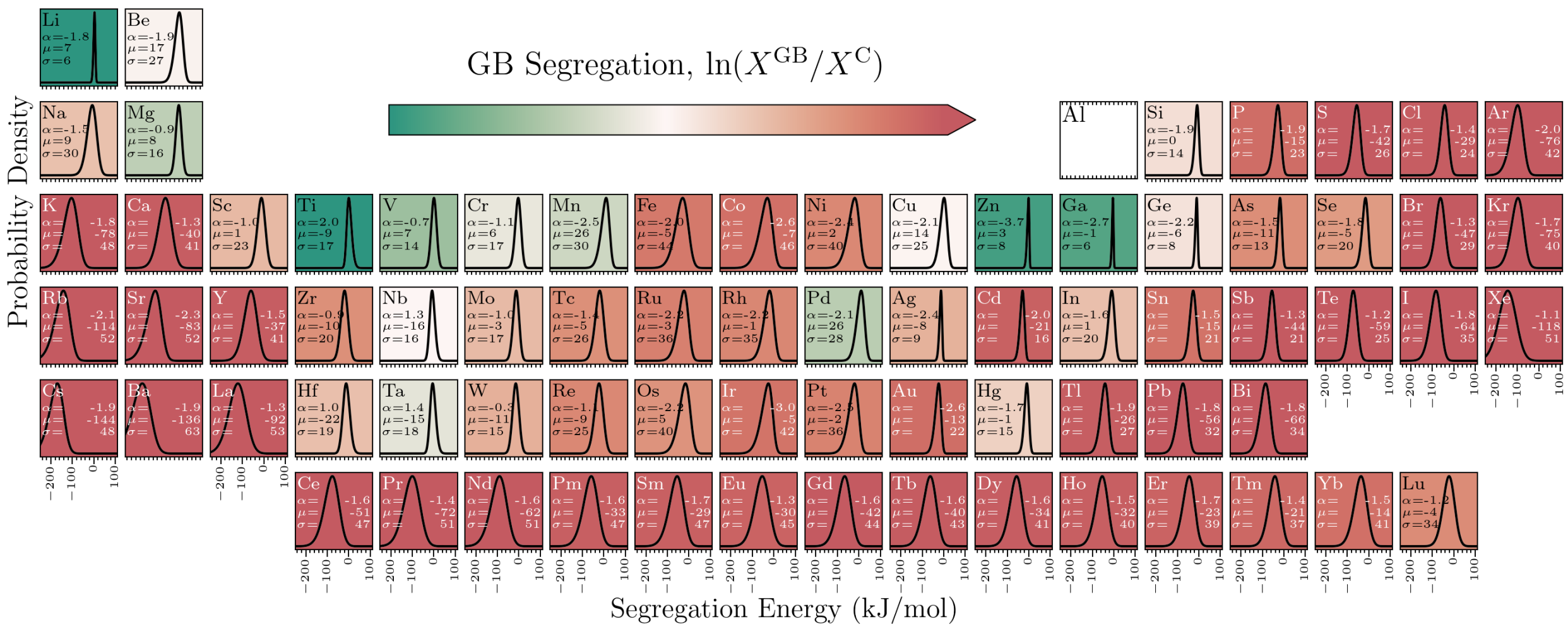

Fig. S3 Segregation spectra for Al-based substitutional binary alloys in kJ/mol

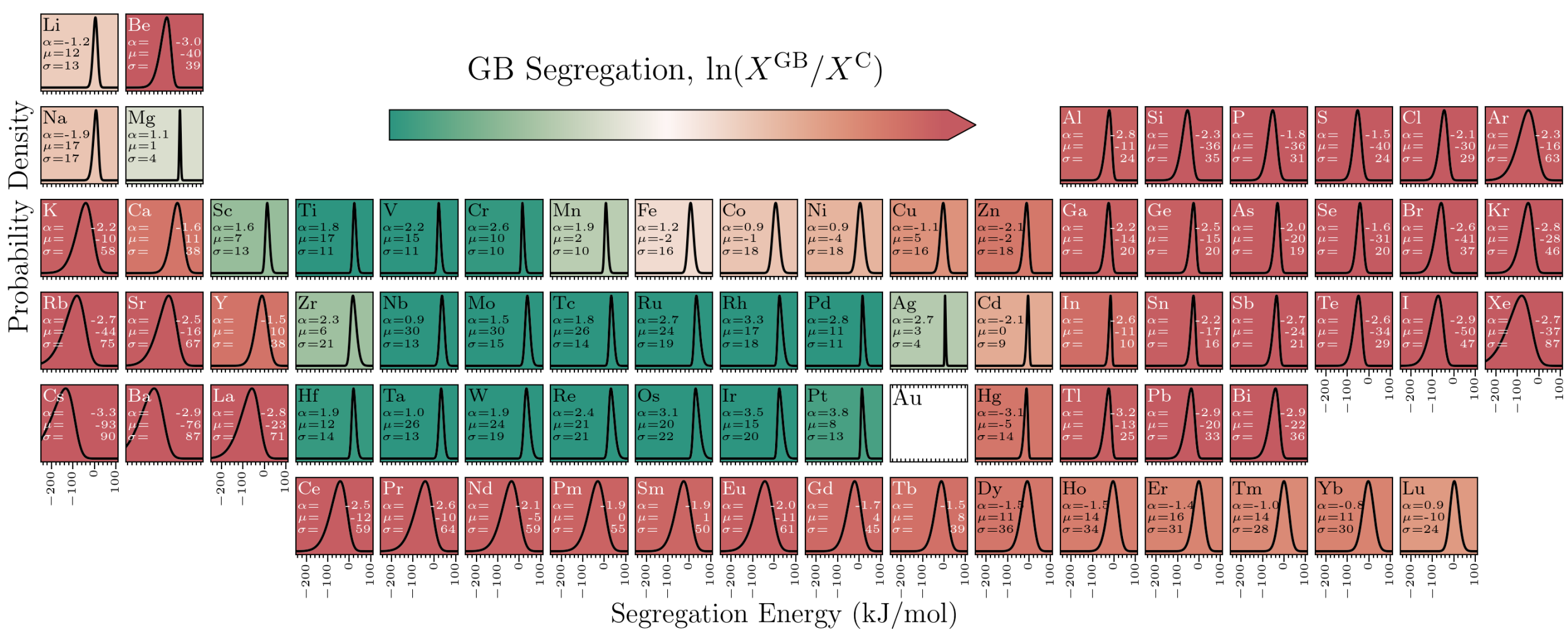

Fig. S4 Segregation spectra for Au-based substitutional binary alloys in kJ/mol

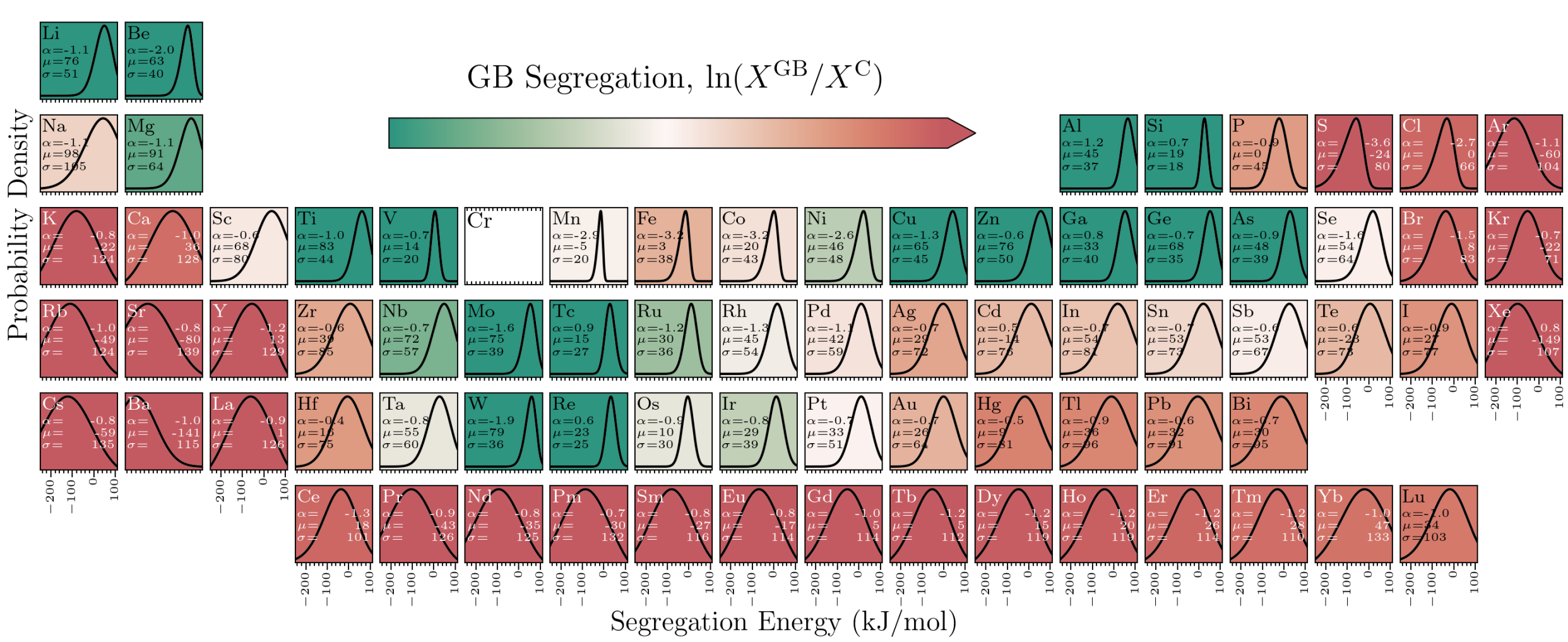

Fig. S5 Segregation spectra for Cr-based substitutional binary alloys in kJ/mol

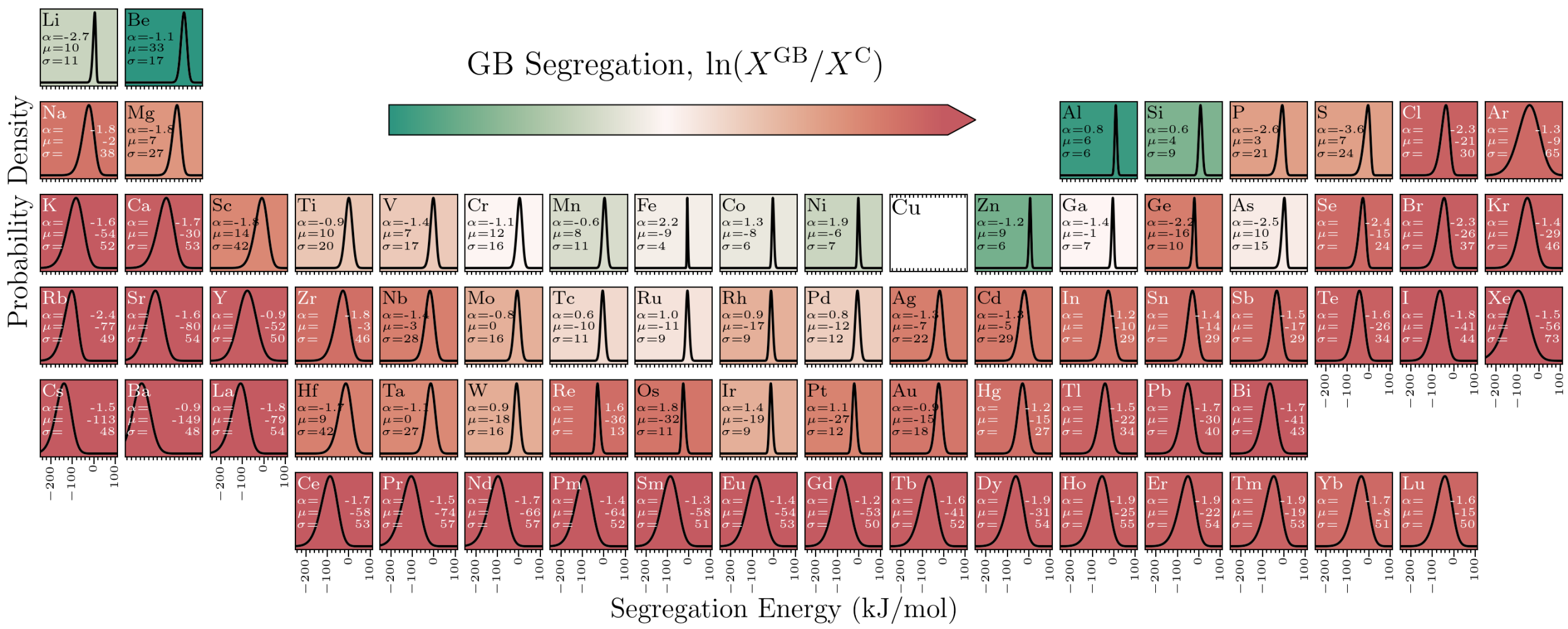

Fig. S6 Segregation spectra for Cu-based substitutional binary alloys in kJ/mol

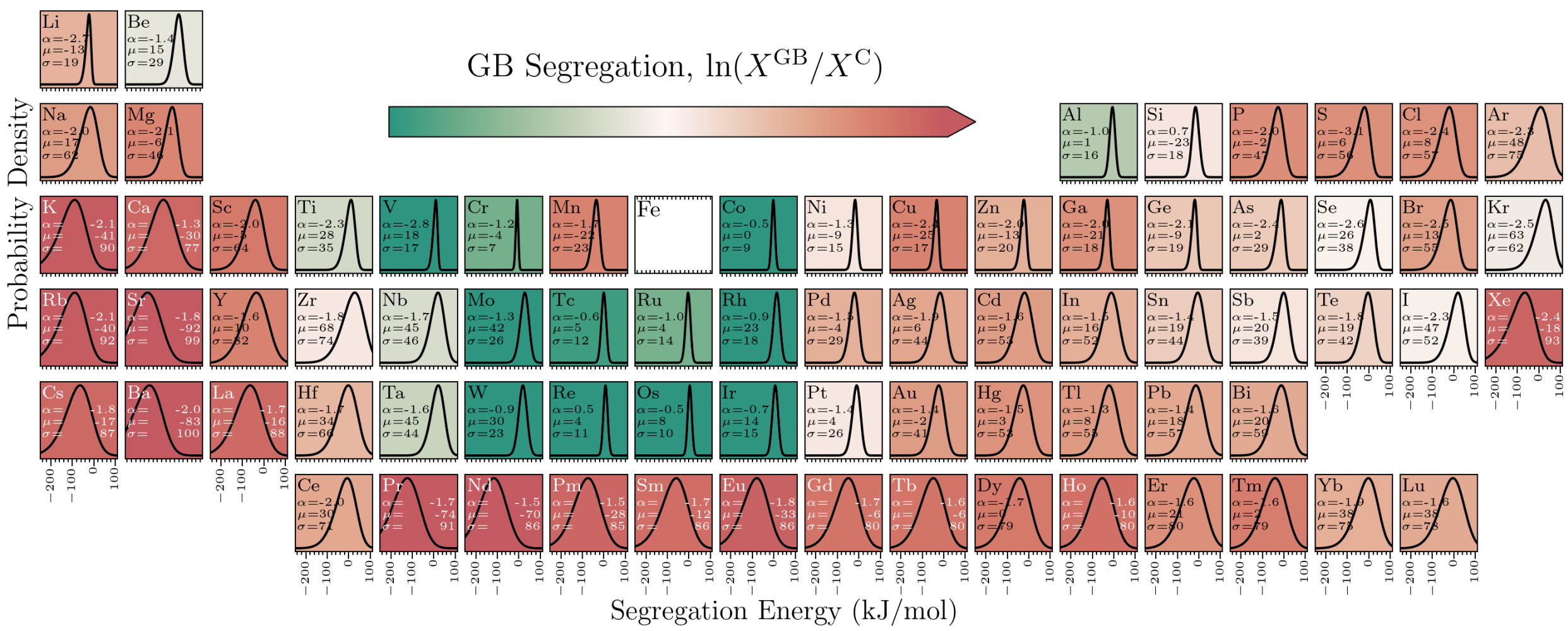

Fig. S7 Segregation spectra for Fe-based substitutional binary alloys in kJ/mol

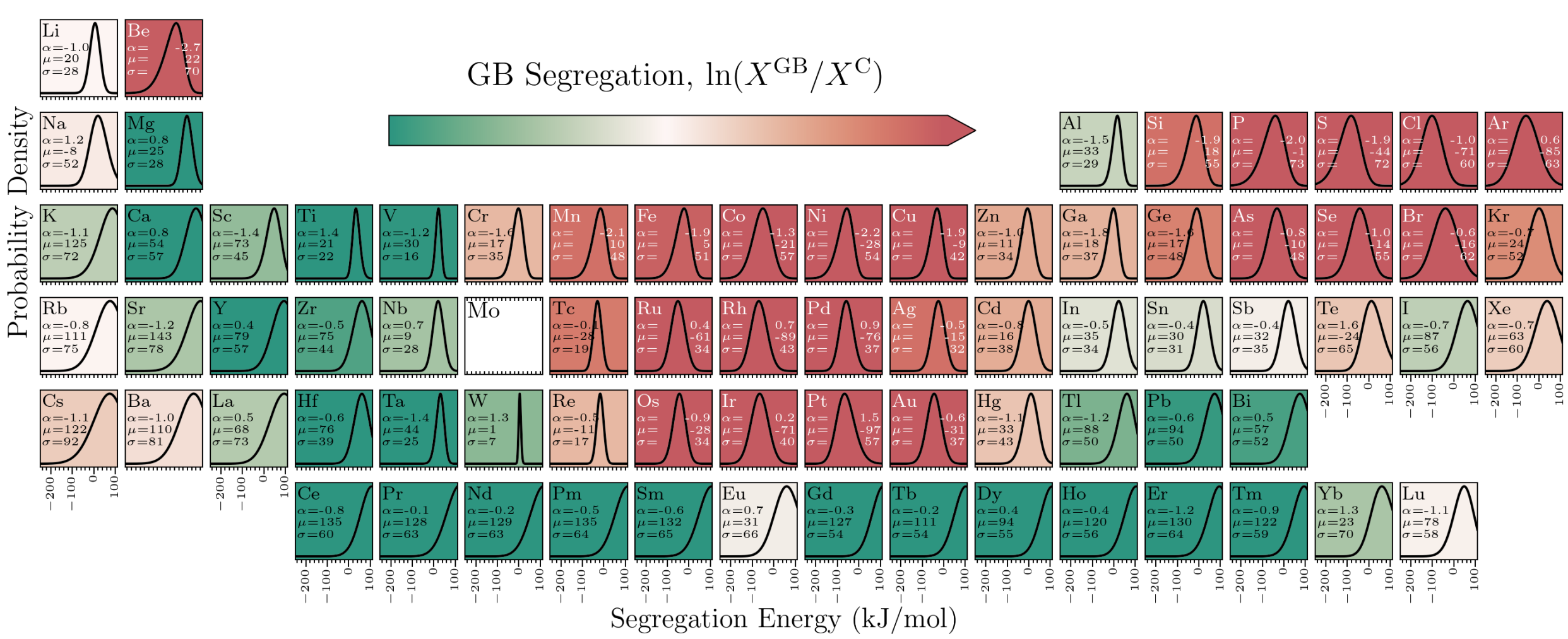

Fig. S8 Segregation spectra for Mo-based substitutional binary alloys in kJ/mol

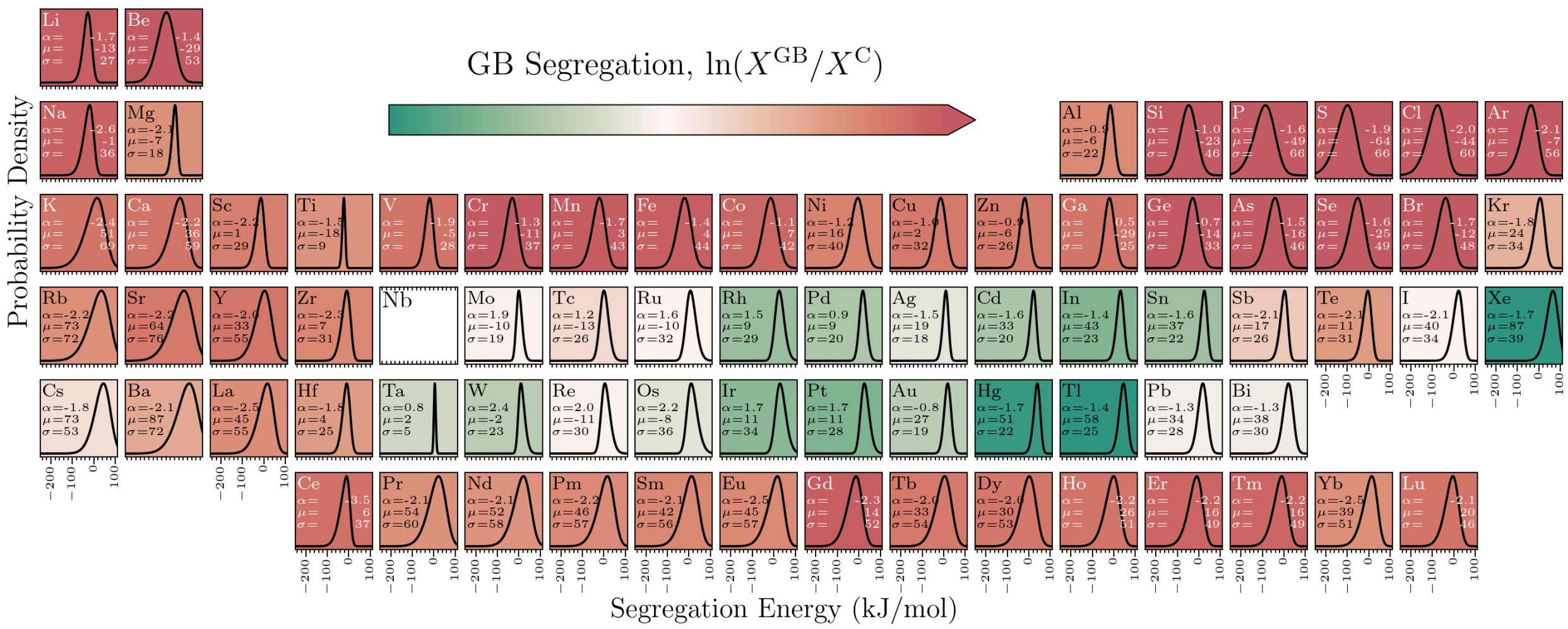

Fig. S9 Segregation spectra for Nb-based substitutional binary alloys in kJ/mol

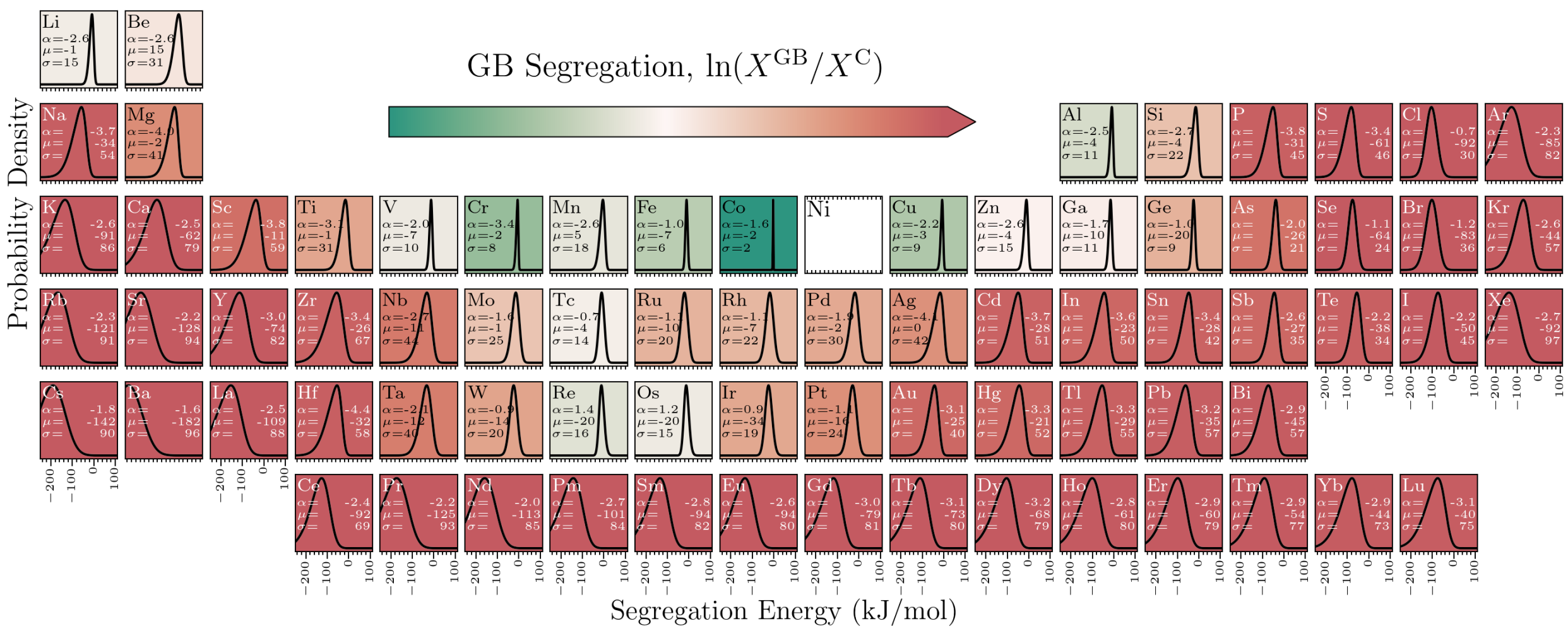

Fig. S10 Segregation spectra for Ni-based substitutional binary alloys in kJ/mol

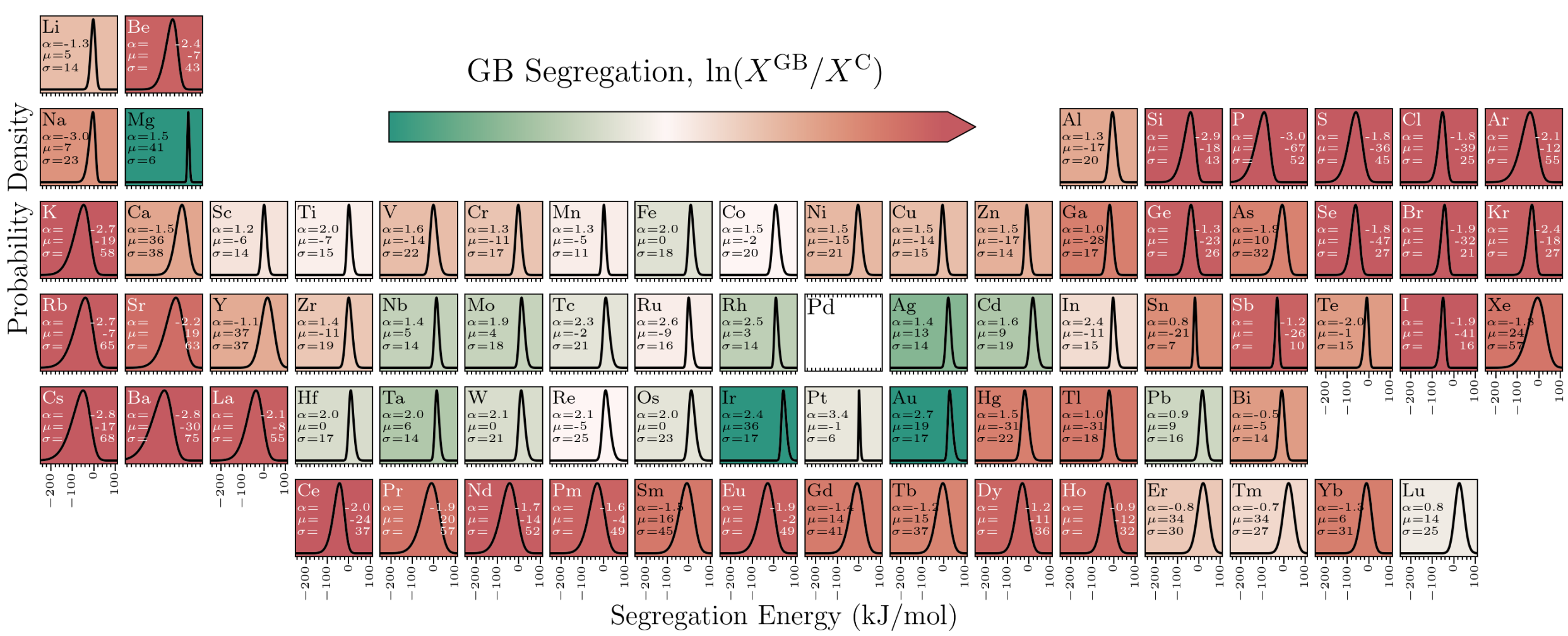

Fig. S11 Segregation spectra for Pd-based substitutional binary alloys in kJ/mol

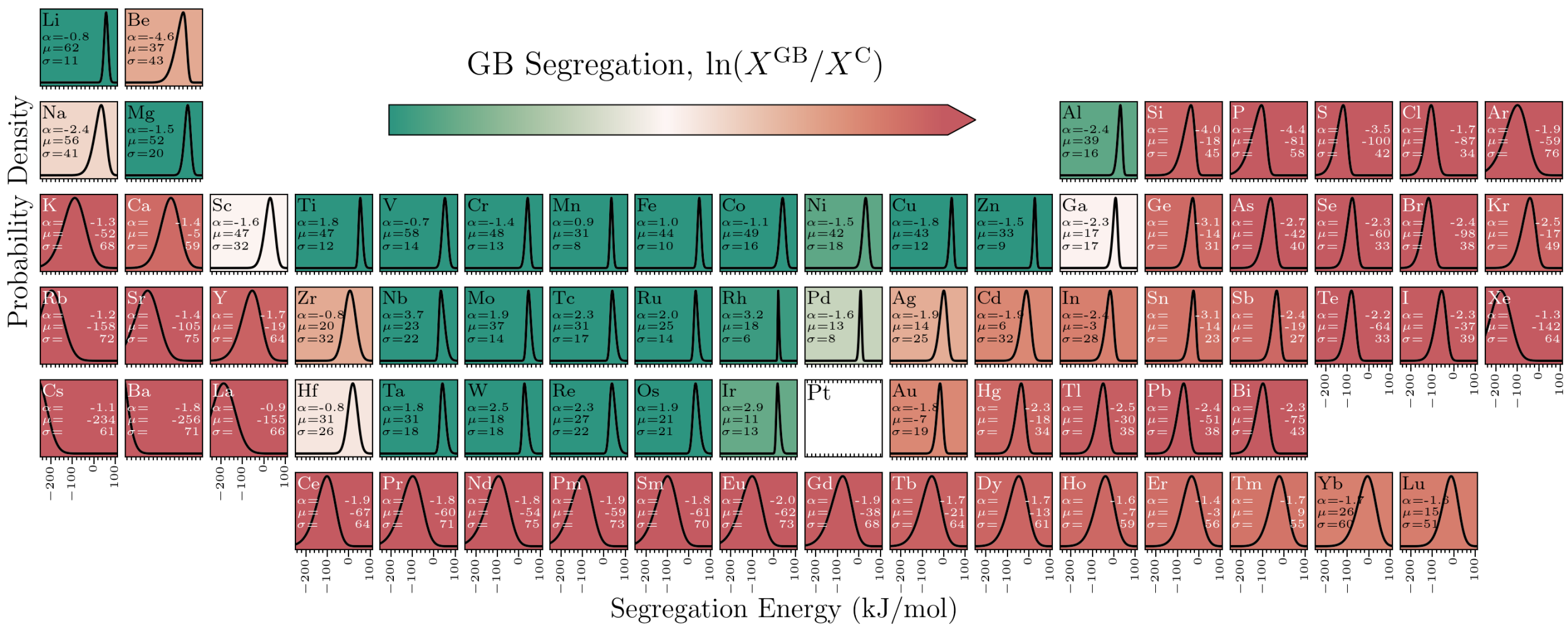

Fig. S12 Segregation spectra for Pt-based substitutional binary alloys in kJ/mol

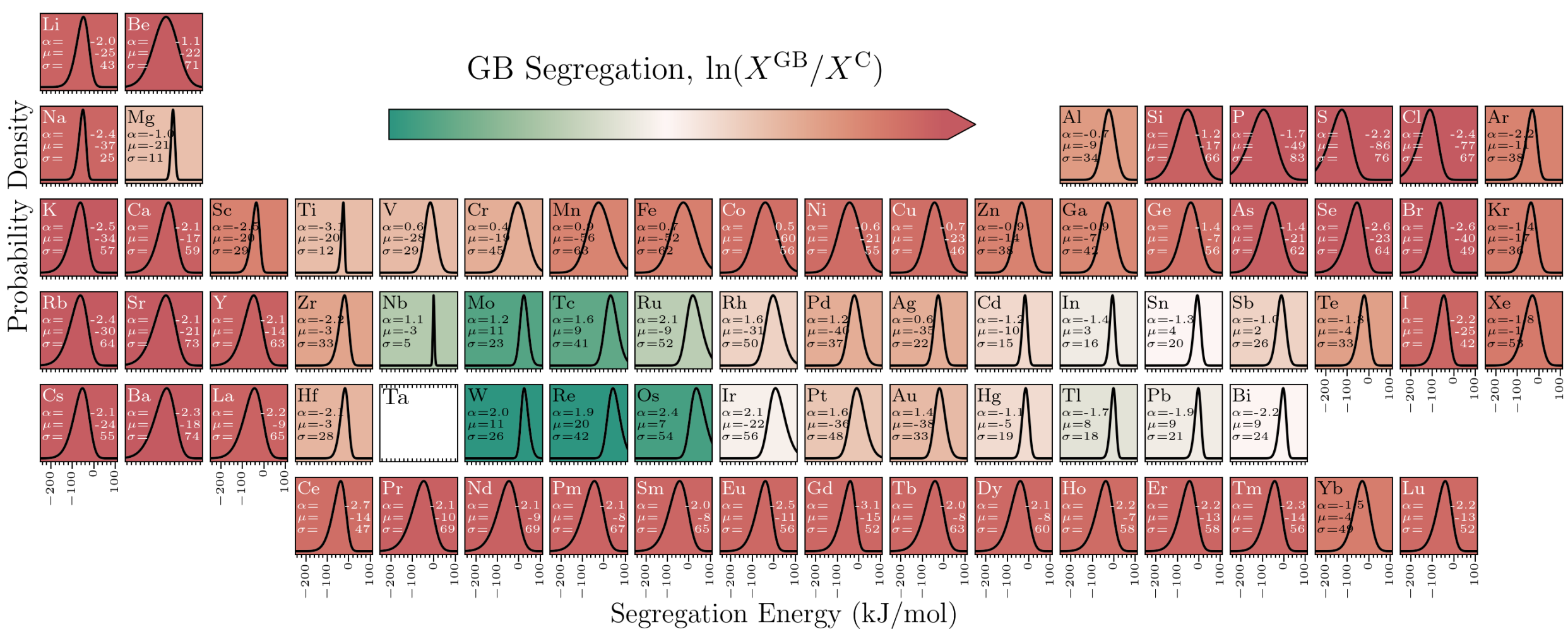

Fig. S13 Segregation spectra for Ta-based substitutional binary alloys in kJ/mol

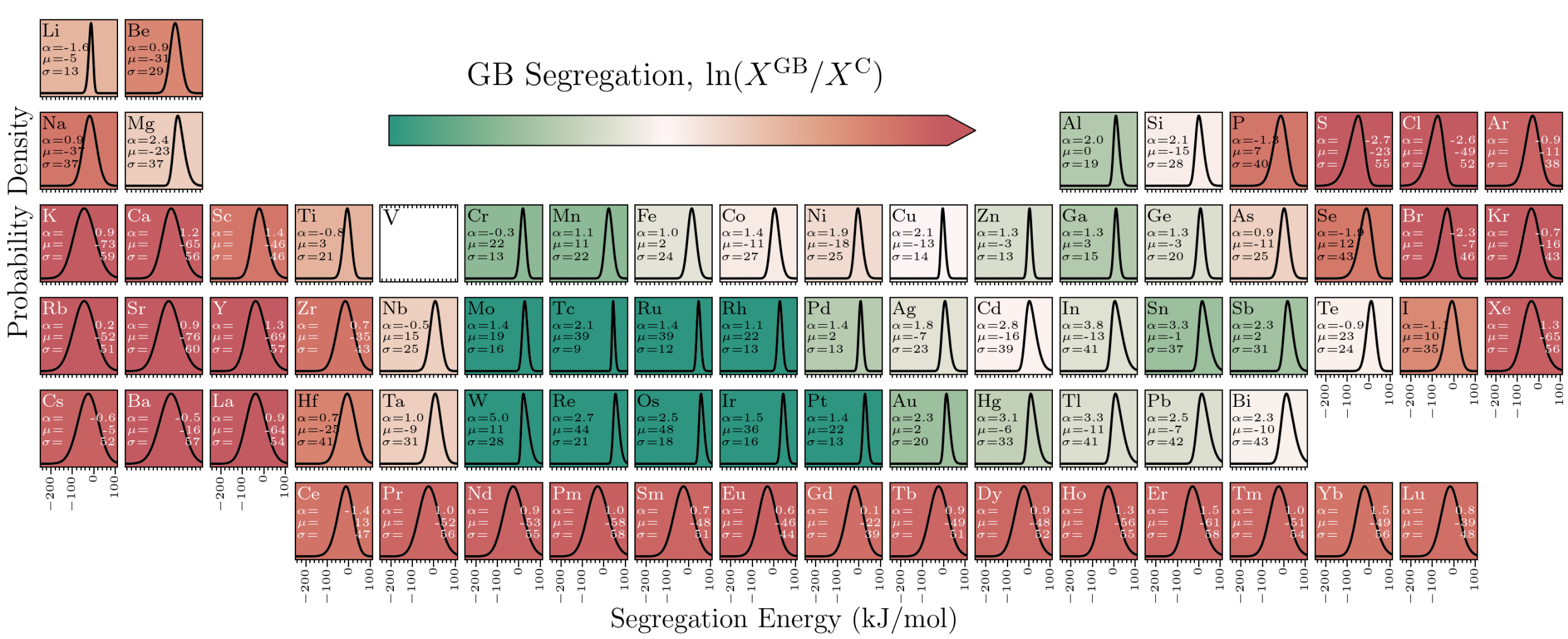

Fig. S14 Segregation spectra for V-based substitutional binary alloys in kJ/mol

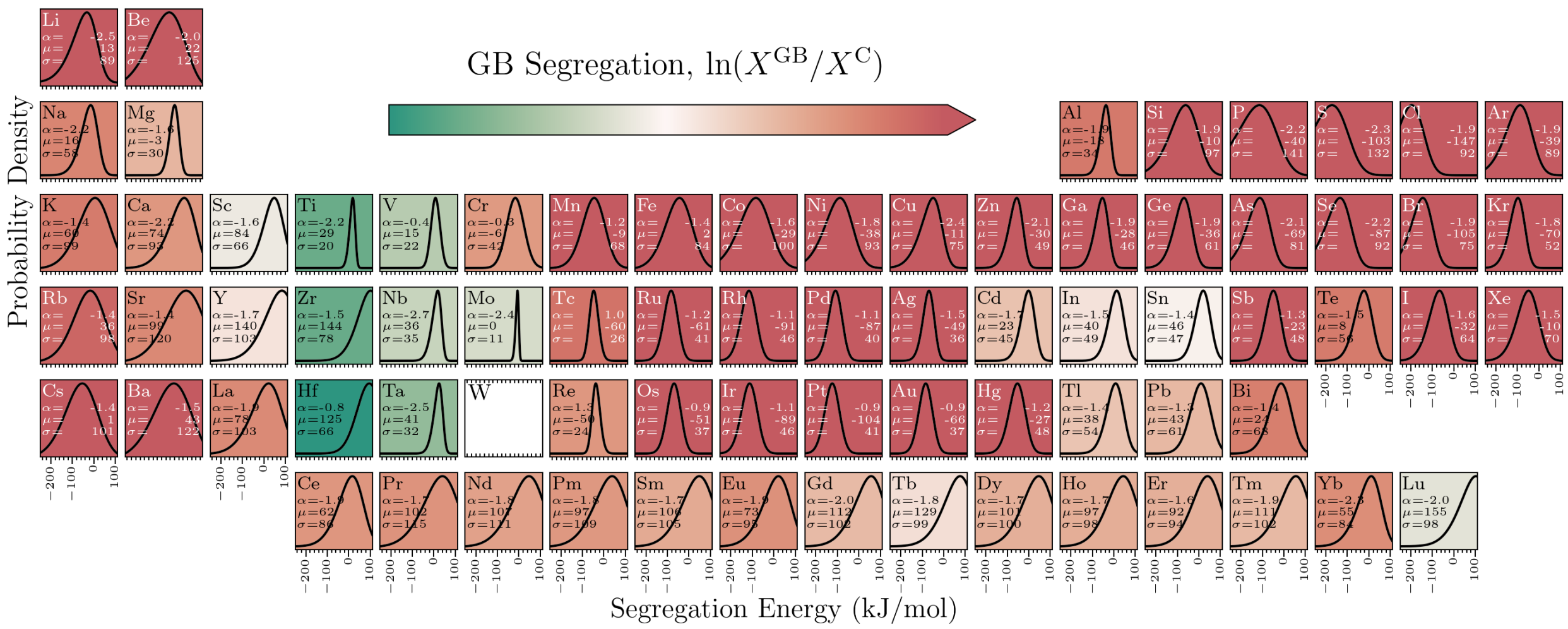

Fig. S15 Segregation spectra for W-based substitutional binary alloys in kJ/mol

## References

[1] A. P. Thompson et al., LAMMPS - a flexible simulation tool for particle-based materials modeling at the atomic, meso, and continuum scales, Computer Physics Communications **271**, 108171 (2022).

[2] W. M. Brown, A. Kohlmeyer, S. J. Plimpton, and A. N. Tharrington, Implementing molecular dynamics on hybrid high performance computers–Particle–particle particle-mesh, Computer Physics Communications **183**, 449 (2012).

[3] W. M. Brown, P. Wang, S. J. Plimpton, and A. N. Tharrington, Implementing molecular dynamics on hybrid high performance computers–short range forces, Computer Physics Communications **182**, 898 (2011).

[4] W. M. Brown and M. Yamada, Implementing molecular dynamics on hybrid high performance computers—three-body potentials, Computer Physics Communications **184**, 2785 (2013).

[5] P. L. Williams, Y. Mishin, and J. C. Hamilton, An embedded-atom potential for the Cu–Ag system, Modelling and Simulation in Materials Science and Engineering **14**, 817 (2006).

[6] M. I. Mendelev, M. J. Kramer, C. A. Becker, and M. Asta, Analysis of semi-empirical interatomic potentials appropriate for simulation of crystalline and liquid Al and Cu, Philosophical Magazine **88**, 1723 (2008).

[7] P. A. T. Olsson, Transverse resonant properties of strained gold nanowires, Journal of Applied Physics **108**, 034318 (2010).

[8] S. M. Eich, D. Beinke, and G. Schmitz, Embedded-atom potential for an accurate thermodynamic description of the iron–chromium system, Computational Materials Science **104**, 185 (2015).

[9] M. I. Mendelev and A. H. King, The interactions of self-interstitials with twin boundaries, Philosophical Magazine **93**, 1268 (2013).

[10] Y. Chen, X. Liao, N. Gao, W. Hu, F. Gao, and H. Deng, Interatomic potentials of W–V and W–Mo binary systems for point defects studies, Journal of Nuclear Materials **531**, 152020 (2020).

[11] M. R. Fellinger, H. Park, and J. W. Wilkins, Force-matched embedded-atom method potential for niobium, Phys. Rev. B **81**, 144119 (2010).

[12] M. I. Mendelev, M. J. Kramer, S. G. Hao, K. M. Ho, and C. Z. Wang, Development of interatomic potentials appropriate for simulation of liquid and glass properties of NiZr2 alloy, Philosophical Magazine **92**, 4454 (2012).

[13] S. M. Foiles, M. I. Baskes, and M. S. Daw, Embedded-atom-method functions for the fcc metals Cu, Ag, Au, Ni, Pd, Pt, and their alloys, Phys. Rev. B **33**, 7983 (1986).

[14] A. Hirata, L. J. Kang, T. Fujita, B. Klumov, K. Matsue, M. Kotani, A. R. Yavari, and M. W. Chen, Geometric frustration of icosahedron in metallic glasses, Science **341**, 376 (2013).

[15] Y. Chen, J. Fang, L. Liu, W. Hu, N. Gao, F. Gao, and H. Deng, Development of the interatomic potentials for W-Ta system, Computational Materials Science **163**, 91 (2019).
[16] M. I. Mendelev, S. Han, W. Son, G. J. Ackland, and D. J. Srolovitz, Simulation of the interaction between Fe impurities and point defects in V, Phys. Rev. B **76**, 214105 (2007).
[17] D. R. Mason, D. Nguyen-Manh, and C. S. Becquart, An empirical potential for simulating vacancy clusters in tungsten, Journal of Physics: Condensed Matter **29**, 505501 (2017).
[18] L. Chanussot, A. Das, S. Goyal, T. Lavril, M. Shuaibi, M. Riviere, K. Tran, J. Heras-Domingo, C. Ho, and W. Hu, Open catalyst 2020 (OC20) dataset and community challenges, Acs Catalysis **11**, 6059 (2021).
[19] L. Barroso-Luque, M. Shuaibi, X. Fu, B. M. Wood, M. Dzamba, M. Gao, A. Rizvi, C. L. Zitnick, and Z. W. Ulissi, Open materials 2024 (omat24) inorganic materials dataset and models, arXiv Preprint arXiv:2410.12771 (2024).
[20] Y.-L. Liao, B. Wood, A. Das, and T. Smidt, Equiformerv2: Improved equivariant transformer for scaling to higher-degree representations, arXiv Preprint arXiv:2306.12059 (2023).
[21] B. Deng, P. Zhong, K. Jun, J. Riebesell, K. Han, C. J. Bartel, and G. Ceder, *CHGNet: Pretrained Universal Neural Network Potential for Charge-Informed Atomistic Modeling*, arXiv:2302.14231.
[22] J. Schmidt, T. F. T. Cerqueira, A. H. Romero, A. Loew, F. Jäger, H.-C. Wang, S. Botti, and M. A. L. Marques, Improving machine-learning models in materials science through large datasets, Materials Today Physics **48**, 101560 (2024).
[23] S. Passaro and C. L. Zitnick, *Reducing SO (3) Convolutions to SO (2) for Efficient Equivariant GNNs*, in *International Conference on Machine Learning* (PMLR, 2023), pp. 27420–27438.
[24] I. Batatia, D. P. Kovacs, G. Simm, C. Ortner, and G. Csányi, MACE: Higher order equivariant message passing neural networks for fast and accurate force fields, Advances in Neural Information Processing Systems **35**, 11423 (2022).
[25] Yuan Chiang, mace-universal, (n.d.).
[26] I. Batatia, S. Batzner, D. P. Kovács, A. Musaelian, G. N. C. Simm, R. Drautz, C. Ortner, B. Kozinsky, and G. Csányi, The design space of E(3)-equivariant atom-centred interatomic potentials, Nature Machine Intelligence **7**, 56 (2025).
[27] I. Batatia, P. Benner, Y. Chiang, A. M. Elena, D. P. Kovács, J. Riebesell, X. R. Advincula, M. Asta, M. Avaylon, and W. J. Baldwin, A foundation model for atomistic materials chemistry, arXiv Preprint arXiv:2401.00096 (2023).
[28] E. Bitzek, P. Koskinen, F. Gähler, M. Moseler, and P. Gumbsch, Structural Relaxation Made Simple, Phys. Rev. Lett. **97**, 170201 (2006).
[29] J. Guénolé, W. G. Nöhring, A. Vaid, F. Houllé, Z. Xie, A. Prakash, and E. Bitzek, Assessment and optimization of the fast inertial relaxation engine (fire) for energy minimization in atomistic simulations and its implementation in lammps, Computational Materials Science **175**, 109584 (2020).
[30] M. Wagih and C. A. Schuh, Learning Grain-Boundary Segregation: From First Principles to Polycrystals, Phys. Rev. Lett. **129**, 046102 (2022).
[31] T. D. Swinburne and J. R. Kermode, Computing energy barriers for rare events from hybrid quantum/classical simulations through the virtual work principle, Phys. Rev. B **96**, 144102 (2017).
[32] N. Tuchinda, G. B. Olson, and C. A. Schuh, Grain Boundary Segregation and Embrittlement of Aluminum Binary Alloys from First Principles, Acta Materialia 121058 (2025).
[33] S. Zhang, O. Y. Kontsevoi, A. J. Freeman, and G. B. Olson, First principles investigation of zinc-induced embrittlement in an aluminum grain boundary, Acta Materialia **59**, 6155 (2011).
[34] R. Hadian, B. Grabowski, and J. Neugebauer, GB code: A grain boundary generation code, The Journal of Open Source Software **3**, (2018).
[35] G. Kresse and J. Hafner, Ab initio molecular dynamics for liquid metals, Phys. Rev. B **47**, 558 (1993).
[36] G. Kresse and J. Furthmüller, Efficiency of ab-initio total energy calculations for metals and semiconductors using a plane-wave basis set, Computational Materials Science **6**, 15 (1996).

[37] G. Kresse and J. Furthmüller, Efficient iterative schemes for ab initio total-energy calculations using a plane-wave basis set, Phys. Rev. B **54**, 11169 (1996).
[38] G. Kresse and D. Joubert, From ultrasoft pseudopotentials to the projector augmented-wave method, Phys. Rev. B **59**, 1758 (1999).
[39] P. E. Blöchl, Projector augmented-wave method, Phys. Rev. B **50**, 17953 (1994).
[40] J. P. Perdew, K. Burke, and M. Ernzerhof, Generalized Gradient Approximation Made Simple, Phys. Rev. Lett. **77**, 3865 (1996).